\documentclass[]{emulateapj}
\usepackage{natbib,amsmath}

\shorttitle{Probing the IGM/Galaxy Connection IV: The LCO/WFCCD
  Galaxy}
\shortauthors{Prochaska et al.}

\begin{document}

\def\nphot{84718}
\def\nspec{2933}
\def\nztot{2800}
\def\nzigm{1198}
\def\wlya{$W^{\rm Ly\alpha}$}
\def\mwlya{W^{\rm Ly\alpha}}
\def\mwovi{W^{\rm OVI}}
\def\wovi{$W^{\rm OVI}$}
\def\mnovi{N(O^{+5})}
\def\novi{$N(O^{+5})$}
\def\nfields{20}
\def\tonso{Ton~S~180}
\def\tonst{Ton~S~210}
\def\mug{(u-g)}
\def\ug{$(u-g)$}
\def\nbin{150}
\def\vbestmfp{48.2 \pm 1.9}
\def\vslopemfp{-37 \pm 4.7}
\def\bestmfp{\lambda_0}
\def\slopemfp{b_\lambda}
\def\omegam{\Omega_{\rm m}}
\def\mnull{\nu_{\rm 912}}
\def\nnull{$\nu_{\rm 912}$}
\def\hub{h_{72}^{-1}}
\def\umfp{{\hub \, \rm Mpc}}
\def\lbr{$\lambda_{\rm r}$}
\def\mlbr{\lambda_{\rm r}}
\def\intl{\int\limits}
\def\kll{$\kappa_{\rm LL}$}
\def\mkll{\kappa_{\rm LL}}
\def\kconst{$\kappa_{\mzq}$}
\def\mkconst{\kappa_{\mzq}}
\def\mztkll{{\tilde\kappa}_{912}}
\def\zq{$z_q$}
\def\mzq{z_q}
\def\maxoff{0.4}
\def\clls{3.50 \pm 0.12}
\def\alls{1.89 \pm 0.39}
\def\blls{-1.0 \pm 0.3}
\def\cmma{\;\;\; ,}
\def\perd{\;\;\; .}
\def\ltk{\left [ \,}
\def\ltp{\left ( \,}
\def\ltb{\left \{ \,}
\def\rtk{\, \right  ] }
\def\rtp{\, \right  ) }
\def\rtb{\, \right \} }
\def\sci#1{{\; \times \; 10^{#1}}}
\def \rAA {\rm \AA}
\def \zem {$z_{\rm em}$}
\def \mzem {z_{\rm em}}
\def \mzlls {z_{\rm LLS}}
\def \zlls {$z_{\rm LLS}$}
\def \mzend {z_{\rm end}}
\def \zend {$z_{\rm end}$}
\def \zstrto {$z_{\rm start}^{\rm S/N=1}$}
\def \mzstrto {z_{\rm start}^{\rm S/N=1}}
\def \zstrt {$z_{\rm start}$}
\def \mzstrt {z_{\rm start}}
\def\smm{\sum\limits}
\def \lll  {$\lambda_{\rm 912}$}
\def \mlll  {\lambda_{\rm 912}}
\def \mtll  {\tau_{\rm 912}}
\def \tll  {$\tau_{\rm 912}$}
\def \avtll  {\tilde{\mtll}}
\def \tigm  {$\tau_{\rm IGM}$}
\def \mtigm  {\tau_{\rm IGM}}
\def \cmm  {cm$^{-2}$}
\def \cmmm {cm$^{-3}$}
\def \kms  {km~s$^{-1}$}
\def \mkms  {{\rm km~s^{-1}}}
\def \lyaf {Ly$\alpha$ forest}
\def \Lya  {Ly$\alpha$}
\def \ovi  {\ion{O}{6}}
\def \mnovi  {N({\rm O^{+5}})}
\def \mlya  {\rm Ly\alpha}
\def \lya  {Ly$\alpha$}
\def \mlya  {Ly\alpha}
\def \Lyb  {Ly$\beta$}
\def \lyb  {Ly$\beta$}
\def \Lyg  {Ly$\gamma$}
\def \lyg  {Ly$\gamma$}
\def \lyd  {Ly$\delta$}
\def \ly5  {Ly-5}
\def \ly6  {Ly-6}
\def \ly7  {Ly-7}
\def \nhi  {$N_{\rm HI}$}
\def \mnhi  {N_{\rm HI}}
\def \lnhi {$\log N_{HI}$}
\def \mlnhi {\log N_{HI}}
\def \etal {\textit{et al.}}
\def \ob {$\Omega_b$}
\def \obh {$\Omega_bh^{-2}$}
\def \om {$\Omega_m$}
\def \ol {$\Omega_{\Lambda}$}
\def \gz {$g(z)$}
\def \mgz {g(z)}
\def \lyaf {Lyman--$\alpha$ forest}
\def \fnhi {$f(\mnhi,z)$}
\def \mfnhi {f(\mnhi,z)}
\def \mfp {$\lambda_{\rm mfp}^{912}$}
\def \mmfp {\lambda_{\rm mfp}^{912}}
\def \btlls {$\beta_{\rm LLS}$}
\def \mbtlls {\beta_{\rm LLS}}
\def \teff {$\tau_{\rm eff,LL}$}
\def \mteff {\tau_{\rm eff,LL}}
\def \O {${\mathcal O}(N,X)$}
\newcommand{\cm}[1]{\, {\rm cm^{#1}}}
\def \zem {$z_{em}$}
\def \snrlim {SNR$_{lim}$}

\title{Probing the IGM/Galaxy Connection IV: The LCO/WFCCD Galaxy Survey of 20
  Fields Surrounding UV Bright Quasars}

\author{
J. Xavier Prochaska\altaffilmark{1}, 
B. Weiner\altaffilmark{2}, 
H.-W. Chen\altaffilmark{3}, 
K.L. Cooksey\altaffilmark{4}, 
J.S. Mulchaey\altaffilmark{5}, 
}
\altaffiltext{1}{Department of Astronomy and Astrophysics \& UCO/Lick
  Observatory, University of California, 1156 High Street, Santa Cruz,
  CA 95064; xavier@ucolick.org}
\altaffiltext{2}{Steward Observatory, University of Arizona, 933
  N. Cherry Ave., Tucson, AZ 85721; bjw@as.arizona.edu}
\altaffiltext{3}{Department of Astronomy; University of Chicago; 5640
  S. Ellis Ave., Chicago, IL 60637; hchen@oddjob.uchicago.edu} 
\altaffiltext{4}{NSF Astronomy \& Astrophysics Postdoctoral Fellow, MIT
  Kavli Institute for Astrophysics \& Space Research, 77 Massachusetts
  Avenue, 37-611, Cambridge, MA 02139, USA; kcooksey@space.mit.edu}
\altaffiltext{5}{Carnegie Observatories; 213 Santa Barbara St.,
  Pasadena, CA 91101; mulchaey@obs.carnegiescience.edu}

\begin{abstract}
We publish the survey for galaxies in \nfields\ fields containing ultraviolet
bright quasars (with $z_{\rm em} \approx 0.1$ to 0.5) that can be used
to study the association between galaxies and absorption systems from the
low-$z$ intergalactic medium (IGM).  The survey is
magnitude limited ($R \approx 19.5$\,mag) and highly complete out to
$10'$ from the quasar in each field.  It was designed to
detect dwarf galaxies ($L \approx 0.1 L^*$) at an impact parameter
$\rho \approx 1$\,Mpc ($z=0.1$) from a quasar.  
The complete sample (all 20 fields) includes $R$-band photometry for 
\nphot\ sources and confirmed redshifts for \nztot\ sources.  This includes 
\nzigm\ galaxies with $0.005 < z < (z_{\rm em} - 0.01)$ at a median
redshift of 0.18, which may associated with IGM absorption lines.
All of the imaging was acquired with cameras on the Swope 40$''$ telescope 
and the spectra were obtained via slitmask observations using the
WFCCD spectrograph on the Dupont 100$''$ telescope at Las
Campanas Observatory (LCO).  This paper describes the data reduction,
imaging analysis, photometry, and spectral analysis of the survey.  We
tabulate the principal measurements for all sources in each field and
provide the spectroscopic dataset online.

\end{abstract}


\section{Introduction}

The intergalactic medium (IGM) is the gas and dust which permeates the
space between galaxies in our universe.  
The IGM is inferred to be the dominant repository of baryons in our
universe; it provides the fuel for galaxies and 
serves as a `sink' for the metals they expel. 
The IGM is almost exclusively
observed through absorption-line techniques in the spectra of distant
objects, e.g.\ quasars, gamma-ray bursts.  The so-called \lya\ forest,
for example, is the thicket of \ion{H}{1} absorption lines which 
trace modest overdensities in the universe.  

Despite its name, which suggests a disconnect from galaxies, the IGM
exhibits several characteristics which connect it to them.  Chief
among these is the presence of metals, generally in the form of
highly-ionized species like \ion{C}{4}, \ion{Si}{4}, and \ion{O}{6}
absorption. The medium is too diffuse and ionized to support {\it in
  situ} star-formation, therefore it is expected that these metals were
produced within galaxies and transported by one or more mechanisms 
to the IGM (e.g.\ galactic-scale winds, tidal stripping).  Identifying the timing, the processes, and
the precise interplay between galaxies and the IGM remains a very
active area of current astrophysical research
\citep[e.g.][]{spa+06,od09,wsd+10}. 

For several decades now, researchers have surveyed the fields
surrounding bright quasars with the goal of connecting galaxies that they
discover to specific characteristics of the IGM (as revealed by rest-frame UV
spectroscopy of the quasars).  These include surveys to explore the
association of galaxies to the \lya\ forest
\citep{mwd+93,lbt+95,stocke+95,ssp96,bowen+02,cpw+05,shone10}; searches for galaxies related
to strong \ion{Mg}{2} lines \citep{bb91,s93,bc09,chg+10}; and
studies of the structures giving rise to \ion{O}{6} absorption
\citep{ts00,bjj+01,savage+02,sts+04,tsg+05,pwc+06,stockeetal06,trippetal06,cpc+08,cm09,wakker09}.
The latter ion has received considerable attention because it has been
proposed to trace the $T \approx 10^{5-7}$\,K phase of the IGM, termed
the warm-hot intergalactic medium (WHIM).  Although the WHIM is a
generic prediction of cosmological simulations
\citep{co99,daveetal01,fangandbryan01},
its low density and modest temperature make it especially elusive to
empirical detection.  The \ion{O}{6} doublet, however, has been
extensively surveyed and owing to its high ionization state may track
a portion of the WHIM.
 
%%%%%%%%%%%%%%%%%%%%%%%%%%%%%%%%%%%%%%%%%%%%%%%%%%%%%%%%%%%%%%%%%%%%%%%%%%%%%%%%%%
%%%%%%%%%%%%%%%%%%%%%%%%%%%%%%%%%%%%%%%%%%%%%%%%%%%%%%%%%%%%%%%%%%%%%%%%%%%%%%%%%%
\begin{deluxetable*}{cccccccccc}[ht]
\tablewidth{0pc}
%\rotate
\tablecaption{Summary of Fields Surveyed\label{tab:fields}}
\tabletypesize{\scriptsize}
\tablehead{\colhead{Quasar} & \colhead{RA} & \colhead{DEC} & 
\colhead{$z_{\rm em}$} & \colhead{{\it HST} UV Spectroscopic Datasets} & 
\colhead{{\it FUSE}$^a$} & \colhead{$R_{\rm max}$} & 
\colhead{$N_{\rm spec}^b$} & \colhead{$C_{19.5}^{5'}$} &
\colhead{$C_{19.5}^{10'}$} \\
& (J2000) & (J2000)
}
\startdata
Q0026+1259 & 00:29:13.8 & +13:16:04       & 0.142 & GHRS/(G270M)& 20&20.0 &  60 & 100 &  91\\
\tonso & 00:57:20.0 & --22:22:56       & 0.062 & STIS/(G140M,G230MB)&132&19.7 &   7 & 100 &  93\\
\tonst & 01:21:51.5 & --28:20:57       & 0.116 & STIS/(E140M,E230M)& 57&20.0 &   7 &  86 &  90\\
PKS0312-77 & 03:11:55.2 & --76:51:51       & 0.223 & STIS/(E140M)&&20.0 &  56 & 100 &  97\\
PKS0405-123 & 04:07:48.4 & --12:11:37       & 0.573 & STIS/(E140M,G230M); GHRS/(G160M,G200M)& 71&20.0 & 565 & 100 &  97\\
PKS0558-504 & 05:59:47.4 & --50:26:52       & 0.137 & STIS/(G230MB)&400&20.0 &  16 & 100 &  99\\
PG1004+130 & 10:07:26.1 & +12:48:56       & 0.240 & STIS/(G140M)& 85&19.5 &  61 &  74 &  71\\
HE1029-140 & 10:31:54.3 & --14:16:51       & 0.086 & STIS/(G140M)&&19.8 &   8 & 100 &  95\\
PG1116+215 & 11:19:08.7 & +21:19:18      & 0.176 & STIS/(G140M,E140M,E230M); GHRS/(G140L)& 76&20.0 &  74 &  76 &  79\\
PG1211+143 & 12:14:17.7 & +14:03:13      & 0.081 & STIS/(G140M,E140M); GHRS/(G140L,G270M)& 52&19.5 &  25 &  61 &  56\\
PG1216+069 & 12:19:20.9 & +06:38:38       & 0.331 & STIS/(E140M); GHRS/(G140L)& 13&20.0 & 101 & 100 &  89\\
3C273 & 12:29:6.70 & +02:03:09      & 0.158 & STIS/(E140M); GHRS/(FG130,FG190,G160M)& 42&20.0 &  32 &  81 &  84\\
Q1230+0947 & 12:33:25.8 & +09:31:23      & 0.415 & GHRS/(G140L)& 13&19.5 &  99 &  83 &  77\\
PKS1302-102 & 13:05:33.0 & --10:33:19       & 0.286 & STIS/(E140M)&140&19.5 &  63 &  89 &  65\\
PG1307+085 & 13:09:47.0 & +08:19:49      & 0.155 & STIS/(G230MB)& 11&19.5 &  41 &  79 &  75\\
MRK1383 & 14:29:06.4 & +01:17:06     & 0.086 & STIS/(E140M,G140M)& 64&19.5 &   5 &  79 &  68\\
Q1553+113 & 15:55:43.0 & +11:11:24      & 0.360 & & 20&20.0 & 106 &  90 &  85\\
PKS2005-489 & 20:09:25.4 & --48:49:54     & 0.071 & STIS/(G140M)& 48&20.0 &  32 &  97 &  97\\
FJ2155-0922 & 21:55:01.5 & --09:22:25     & 0.192 & STIS/(E140M,G230MB)& 46&20.0 & 105 & 100 &  95\\
PKS2155-304 & 21:58:51.8 & --30:13:30     & 0.116 & STIS/(E140M); GHRS/(G160M,ECH-B,G140L)&123&20.0 &  43 &  96 &  96\\
\enddata
\tablecomments{Columns 9 and 10 list the completeness percent of the spectroscopic survey in each field to $R=19.5$\,mag for a radius of 5 and 10 arcminutes respectively.}
\tablenotetext{a}{Total integration time in ks.}
\tablenotetext{b}{Number of spectroscopically determined galaxy redshifts for objects with $0.005 < z < z_{\rm em}$.}
\end{deluxetable*}

Motivated by the analysis of \ion{O}{6} and its relation to
galaxies and the WHIM,  we initiated a program to survey galaxies in
the fields surrounding UV-bright, $z_{\rm em} \gtrsim 0.1$ quasars that were
targeted (or likely to be targeted) by the {\it Far Ultraviolet
  Spectroscopic Explorer (FUSE)} and/or the {\it Hubble Space
  Telescope (HST)}.
Prior to the commissioning
of the Cosmic Origins Spectrograph (COS), there were only a small
set of known quasars that were bright enough for such observations.  
The sources are widely separated across the sky such that
existing, ongoing, and/or planned galaxy surveys (e.g.\ SDSS, 2dF)
do not cover the fields or are often too shallow for studying $L >
L^*$ galaxies beyond $z \approx 0.05$.
Therefore, we designed a survey
with two main observational goals: 
(1) achieve high completeness for galaxies
as faint as $\approx 0.1 L^*$ at $z=0.1$; and
(2) survey an area corresponding to at least 1\,Mpc at $z=0.1$.
Technically, this implied a magnitude limit of $R \approx 19.5$\,mag
and a field-of-view (FOV) of roughly $10'$ radius around each quasar.
These requirements were well-suited to the WFCCD
spectrograph on the $100''$ Dupont Telescope at Las Campanas
Observatory (LCO).  Over the course of several years, we targeted \nfields\
fields at equatorial and Southern declinations
(Table~\ref{tab:fields}).
Early results from this survey have been used to explore the
IGM/galaxy connection for \ion{O}{6} absorbers
\citep{pwc+06,cpc+08},
the \lya\ forest \citep{cpw+05}, and a Lyman limit absorber
\citep{lpk+09}. 

With this paper, we provide the full data release of the LCO/WFCCD survey.  This
stands, for now, as the largest dataset for exploring the association
between the IGM and galaxies and their large-scale structures at $z
\sim 0.1$.  
This manuscript comments on galaxies derived only from our own
LCO/WFCCD survey and we emphasize that other datasets (e.g.\ SDSS) and
publications do provide complementary coverage for about half of the
fields.
In future (and previous) papers,
we analyze the IGM/galaxy connection with this dataset.
We organize the paper as follows: $\S$~\ref{sec:obs} describes the 
observations, data reduction, and spectral analysis for redshift
determinations; we present our survey of the fields in
$\S$~\ref{sec:fields}; and offer a brief summary in
$\S$~\ref{sec:summary}.
Unless otherwise specified, we adopt the 5yr WMAP cosmology 
\citep{wmap05}:
$\Omega_\Lambda=0.74,
\Omega_m=0.26, H_0=72 \rm \, km \, s^{-1} \, Mpc^{-1}$. 
Furthermore, all distances are quoted in proper units
unless otherwise denoted.

\begin{deluxetable*}{cccccl}
\tablewidth{0pc}
\tablecaption{Log of Imaging Observations \label{tab:imaging}}
\tabletypesize{\scriptsize}
\tablehead{\colhead{Field} & \colhead{UT Date} & 
 \colhead{CCD} & \colhead{Filter} & 
 \colhead{Exp.} & \colhead{Conditions} 
} 
\startdata
Q0026+1259    & 26 Oct 2000 & SITe1 & B & 2x450s & Photometric \\
              &             &       & R & 3x600s & Photometric \\
Q0046+112     & 26 Oct 2000 & SITe1 & B & 2x450s & Photometric \\
              &             &       & R & 3x600s & Photometric \\
% 00 57
Ton S 180     & 26 Oct 2000 & SITe1 & B & 2x450s & Photometric \\
              &             &       & R & 3x600s & Photometric \\
HE0153$-$4520 & 14 Sep 2001 & SITe3 & B & 5x450s & Photometric \\
              &             &       & R & 6x600s & Photometric \\
% 01 21
Ton S 210     & 26 Oct 2000 & SITe1 & B & 2x450s & Photometric \\
              &             &       & R & 3x600s & Photometric \\
              & 02 Oct 2002 & SITe3 & R &12x600s & Cirrus \\
              & 03 Oct 2002 & SITe3 & R & 9x600s & \\
              & 04 Oct 2002 & SITe3 & R &13x600s & \\
PKS0312$-$770 & 02 Oct 2002 & SITe3 & R & 6x450s & Cirrus \\
              & 04 Oct 2002 & SITe3 & R & 3x600s &  \\
PKS0405$-$123 & 26 Oct 2000 & SITe1 & B & 2x450s & Windy, photometric \\
              &             &       & R & 5x600s & Windy, photometric \\
              & 03 Oct 2002 & SITe3 & R &18x600s &  \\
              & 04 Oct 2002 & SITe3 & R &16x600s &  \\
HE0450$-$2958 & 26 Feb 2001 & SITe1 & B & 2x450s & Photometric \\
              &             &       & R & 3x600s & Photometric \\
% 0513-002
AKN 120       & 23 Feb 2001 & SITe1 & B & 2x450s & Photometric \\
              &             &       & R & 4x600s & Photometric \\
PKS0558$-$504 & 26 Oct 2000 & SITe1 & B & 2x450s & Windy, photometric \\
              &             &       & R & 4x600s & Windy, photometric \\
B0736+017     & 21 Feb 2001 & SITe1 & B & 2x450s & Photometric \\
              &             &       & R & 4x600s & Photometric \\
PG0832+25     & 26 Feb 2001 & SITe1 & B & 2x450s & Photometric \\
              &             &       & R & 3x600s & Photometric \\
IR0914$-$6206 & 21 Feb 2001 & SITe1 & B & 2x450s & Photometric \\
              &             &       & R & 3x600s & Photometric \\
PG1001+054    & 24 Feb 2001 & SITe1 & B & 2x450s & Photometric \\
              &             &       & R & 4x600s & Photometric \\
PG1004+130    & 24 Feb 2001 & SITe1 & B & 2x500s & Photometric \\
              &             &       & R & 4x600s & Photometric \\
PG1011$-$040  & 24 Feb 2001 & SITe1 & B & 2x450s & Photometric \\
              &             &       & R & 3x600s & Photometric \\
HE1015$-$1618 & 27 Feb 2001 & SITe1 & B & 2x450s & Photometric \\
              &             &       & R & 3x600s & Photometric \\
HE1029$-$140  & 23 Feb 2001 & SITe1 & B & 2x500s & Photometric \\
              &             &       & R & 3x600s & Photometric \\
              & 08 Apr 2003 & SITe3 & B &13x600s & Cirrus \\
HE1050$-$2711 & 28 Feb 2001 & SITe1 & B & 2x450s & Photometric \\
              &             &       & R & 3x600s & Photometric \\
HE1115$-$1735 & 27 Feb 2001 & SITe1 & B & 2x450s & Photometric \\
              &             &       & R & 3x600s & Photometric \\
PG1116+215    & 21 Feb 2001 & SITe1 & B & 3x600s & Photometric \\
              &             &       & R & 6x600s & Photometric \\
              & 23 Feb 2001 &       & B & 3x600s & Photometric \\
              &             &       & R & 3x600s & Photometric \\
HE1122$-$1649 & 23 Feb 2001 & SITe1 & B & 2x500s & Photometric \\
              &             &       & R & 3x600s & Photometric \\
% 1126-0
MRK 1298      & 07 Apr 2002 & SITe3 & B & 6x600s & Photometric \\
              &             &       & R &12x600s & Photometric \\
PG1126+215    & 25 Feb 2001 & SITe1 & R & 7x600s & Photometric \\
PG1211+143    & 09 Apr 2002 & SITe3 & B & 6x600s & Photometric \\
              &             &       & R &14x600s & Photometric \\
PG1216+069    & 25 Feb 2001 & SITe1 & B & 2x600s & Photometric \\
              &             &       & R & 4x600s & Photometric \\
              & 26 Feb 2001 &       & B & 4x600s & Photometric \\
              &             &       & R & 7x600s & Photometric \\
\enddata
%\tablecomments{Spline points for evaluating \fnhi\ at $z=3.7$.}
%\tablenotetext{a}{Approximate field of view of the combined images.}
\tablecomments{[The complete version of this table is in the electronic edition of the Journal.  The printed edition contains only a sample.]}
\end{deluxetable*}

\begin{deluxetable*}{ccccl}
\tablewidth{0pc}
\tablecaption{Log of WFCCD Observations \label{tab:wfccd}}
\tabletypesize{\footnotesize}
\tablehead{\colhead{Field} & \colhead{UT Date} & 
 \colhead{Mask} & 
 \colhead{Exp.} & \colhead{Conditions} 
} 
\startdata
Q0026+1259    & 31 Oct 2002 & M1 & 3600 & Clear \\
              & 01 Nov 2002 & M2 & 3600 & Clear\\
              & 02 Nov 2002 & M3 & 3600 & Clear\\
              & 03 Nov 2002 & M4 & 3600 & Clear\\
              & 04 Nov 2002 & M5 & 3600 & Clear\\
              & 05 Nov 2002 & M6 & 1800 & Clear\\
              & 06 Nov 2002 & M6 & 1800 & Clear\\
Ton S 180     & 10 Sep 2001 & M1 & 5400 & Photometric \\
              & 12 Sep 2001 & M2 & 7200 & Clear, poor seeing\\
              & 14 Sep 2001 & M3 & 3600 & Clear\\
              & 15 Sep 2001 & M4 & 3600 & Clear\\
              & 16 Sep 2001 & M5 & 3600 & Clear\\
Ton S 210     & 12 Sep 2001 & M1 & 5400 & Cirrus \\
              & 13 Sep 2001 & M2 & 3600 & Clear \\
              & 13 Sep 2001 & M4 & 3600 & Clear \\
              & 14 Sep 2001 & M3 & 3600 & Clear\\
PKS0312$-$770 & 01 Nov 2002 & M1 & 3600 & Clear \\
              & 01 Nov 2002 & M2 & 3600 & Clear\\
              & 02 Nov 2002 & M3 & 3000 & Clear\\
              & 03 Nov 2002 & M4 & 3300 & Clear\\
              & 04 Nov 2002 & M5 & 3600 & Clear\\
PKS0405$-$123 & 13 Sep 2001 & M1 & 3600 & Cirrus \\
              & 14 Sep 2001 & M3 & 3600 & Clear\\
              & 15 Sep 2001 & M4 & 3600 & Clear\\
              & 16 Sep 2001 & M2 & 3600 & Clear\\
              & 31 Oct 2002 & M5 & 3600 & Clear\\
              & 31 Oct 2002 &  S & 1800 & Clear\\
              & 02 Nov 2002 & EE1& 3600 & Clear\\
              & 02 Nov 2002 & EE2& 3600 & Clear\\
              & 02 Nov 2002 & EE3& 3000 & Clear\\
              & 03 Nov 2002 & NN1& 3000 & Clear\\
              & 03 Nov 2002 & NN2& 3000 & Clear\\
              & 03 Nov 2002 & NN3& 3000 & Clear\\
              & 03 Nov 2002 & NN4& 3300 & Clear\\
              & 04 Nov 2002 & WW1& 3600 & Clear\\
              & 04 Nov 2002 & WW2& 3600 & Clear\\
              & 04 Nov 2002 & WW3& 3600 & Clear\\
              & 04 Nov 2002 & WW4& 3600 & Clear\\
              & 05 Nov 2002 & SS1& 4200 & Clear\\
              & 05 Nov 2002 & SS2& 3600 & Clear\\
              & 05 Nov 2002 & SS3& 3600 & Clear\\
              & 05 Nov 2002 & SS4& 3600 & Clear\\
PKS0558$-$504 & 13 Sep 2001 & M1 & 3600 & Cirrus \\
              & 14 Sep 2001 & M3 & 3600 & Clear\\
              & 15 Sep 2001 & M4 & 3600 & Clear\\
              & 16 Sep 2001 & M2 & 3600 & Clear\\
              & 31 Oct 2002 & M5 & 4200 & Clear\\
              & 01 Nov 2002 & M6 & 3600 & Clear\\
              & 01 Nov 2002 & M7 & 1800 & Clear\\
              & 02 Nov 2002 & M7 & 1800 & Clear\\
              & 06 Nov 2002 & M2 & 1800 & Clear\\
PG1001+054    & 22 Apr 2001 & S0 & 3600 & Photometric \\
PG1004+130    & 31 May 2003 & M1 & 3600 & Cirrus \\
              & 01 Jun 2003 & M2 & 3600 & Cirrus \\
              & 02 Jun 2003 & M3 & 3600 & Clear \\
HE1029$-$140  & 18 Apr 2001 & S0 & 3600 & Photometric \\
              & 19 Apr 2001 & B0 & 5400 & Photometric \\
              & 20 Apr 2001 & B1 & 3600 & Light cirrus \\
              & 21 Apr 2001 & S1 & 3600 & Cirrus \\
PG1116+215    & 18 Apr 2001 & B0 & 3600 & Photometric \\
              & 18 Apr 2001 & B1 & 3600 & Photometric \\
              & 19 Apr 2001 & S0 & 5400 & Photometric \\
              & 20 Apr 2001 & B2 & 3600 & Light cirrus \\
              & 22 Apr 2001 & F1 & 3600 & Photometric \\
              & 05 Jun 2003 & F1 & 3600 & Clouds \\
              & 06 Jun 2003 & F1 & 5400 & Clear \\
PG1211+143    & 31 May 2003 & M2 & 3600 & Cirrus \\
              & 01 Jun 2003 & M3 & 3600 & Cirrus \\
              & 02 Jun 2003 & M1 & 3600 & Clear \\
PG1216+069    & 03 Jun 2003 & M1 & 5400 & Clear \\
              & 03 Jun 2003 & M2 & 5150 & Cirrus \\
              & 03 Jun 2003 & M3 & 3600 & Clear \\
              & 03 Jun 2003 & M4 & 3600 & Clear \\
              & 06 Jun 2003 & M2 & 3600 & Clear \\
\enddata
%\tablecomments{Spline points for evaluating \fnhi\ at $z=3.7$.}
%\tablenotetext{a}{Approximate field of view of the combined images.}
\tablecomments{[The complete version of this table is in the electronic edition of the Journal.  The printed edition contains only a sample.]}
\end{deluxetable*}

\section{Observations and Data Reduction}
\label{sec:obs}

We selected 20 fields accessible from Las Campanas Observatory
(declination $\delta < 25$\,deg) surrounding UV bright quasars with
emission redshifts $z_{\rm em} > 0.05$.  These quasars had
existing UV spectroscopy from the {\it HST} or {\it FUSE} telescopes,
had planned UV spectroscopic observations with one of these
telescopes, and/or have sufficiently
high UV flux to permit UV spectroscopic observations with one of
these facilities.  The quasars were also chosen to have a wide range of right
ascension, to enable a year-round observing campaign.
The quasars and their UV spectroscopic datasets
are summarized in Table~\ref{tab:fields}.

All of the fields in our survey were imaged with the
Swope~$40''$ telescope using a direct imaging camera.  The majority
were taken with the SITe1 CCD which has $0.6964''$ pixels in a $2048
\times 2048$ array.  The remainder were observed with the SITe3 CCD
which has $0.435''$ pixels in a $2048 \times 3150$ array.  All of the
fields were imaged through an $R$-band filter and most also had
contemporaneous $B$-band images taken.  The data were obtained in a series of
dithered exposures intended to map at least a $20' \times 20'$
field-of-view (FOV).  This was designed to enable a survey 
to approximately a 1\,Mpc impact parameter $\rho$ at $z \approx 0.1$.  
Most of the imaging data were
obtained under photometric conditions; 
Table~\ref{tab:imaging} summarizes the weather and provides a log of
the observations.  Most of the data were acquired in good seeing
conditions, i.e.\ FWHM~$\approx 1''$.

The imaging data were reduced with standard IRAF routines to overscan
subtract and flat-field the images.  We used custom routines to
determine integer offsets between the dithered images and combined
the frames after weighting by a global measure of the inverse variance.  
We derived a photometric solution for each night from analysis of
Landoldt standard stars \citep{lan92} that were observed at a range of
airmass.  We solved for the zeropoint, airmass and $(B-R)$ color
terms.  We estimate the systematic uncertainty related to the
photometric solution to be approximately $0.05$\,mag. 

Each combined $R$-band image was analyzed with the SExtractor
\citep[v2.0][]{bertin96}
package.  The parameters were set to require a minimum detection area
of 6~pixels and a detection threshold of $1.5\sigma$ above the RMS of
the sky background.   For each object detected, SExtractor reports a
star-galaxy (S/G) classifier with values ranging from 0 to 1 where unity
indicates a point-spread-function consistent with a point
source (i.e.\ a likely star). Finally, we produced a segmentation map of each
field and calculated $B$ and $R$ magnitudes by summing the flux
in the pixels assigned to each object.  There are no corrections made
for flux that falls below our detection threshold.  The
true galaxy magnitudes, therefore, may be up to a few tenths
magnitude brighter especially for low-surface brightness galaxies.

Proper astrometry for each field was applied with the publicly-
available Image World Coordinate Setting Program
\emph{imwcs}\footnote{http://tdc-www.harvard.edu/software/wcstools/imwcs/index.html.}
version 3.6.8. First, several World Coordinate System (WCS) keywords
were added to
the reduced and combined image header in order to define its estimated
pointing and plate scale ($0.6964''\,{\rm pix}^{-1}$ for the SITe1 CCD and
$0.4349''\,{\rm pix}^{-1}$ for SITe3). Next, we used \emph{imwcs} to
fit a plane-tangent projection, using stars in the image that
\emph{imwcs} identified
and stars from a reference catalog, namely, the US Naval Observatory
astrometric catalog
USNO-A2.0.\footnote{http://tdc-www.harvard.edu/software/catalogs/ua2.html.}
The \emph{imwcs} algorithm typically matched a dozen or more stars and
converged in six
iterations, with a mean arcsec offset of $0.15''$ in any given image.
Finally, all astrometized images were visually compared to the USNO-A2.0 sample
to verify that the solution held across the whole field (e.g., no systematic
rotation was obvious towards the outskirts of the image).

Based on our SExtractor analysis of the images, we designed a set of
slit masks for each field to be used with the WFCCD instrument on the
Dupont $100''$ telescope.  Targets were identified as
follows.  First, we examined the isophotal area versus $R$ magnitude
from the imaging of a given field.  In all cases, stars trace out a
well-defined locus for $R<19.5$\,mag \citep[see Fig.~1 of][for an
example]{pwc+06}.  We traced this locus 
and discarded all objects that fall within the stellar region or fall
just outside the locus and have a S/G value greater than 0.98.
We also did not target the few, very bright ($R<15$\,mag) and very
large (angular diameter $>10''$) galaxies that exist in the survey
fields.  Most of these have previously published redshifts and/or lie
at too low redshift ($z<0.01$) for our main scientific program.
For those that remained, we targeted all objects
with $R \le R_{\rm max}$  and
$\theta < \theta_{\rm max}$ of the QSO.  We set
$R_{\rm max} \ge 19.5$\,mag, adopting fainter values for fields with a
lower surface density of targets on the sky or where we had particular
scientific interest in the field (see Table~\ref{tab:fields}).
For all but two of the fields, $\theta_{\rm max} = 10'$.  The two
exceptions are PKS0312$-$770 and PKS0405$-$123, where we expanded the
survey to cover $\theta_{\rm max} \approx 15'$ and $\approx 20'$
respectively.

We performed multi-slit spectroscopy on each field with the WFCCD
spectrometer using the blue grism 
and the Tek\#5 CCD.  This affords a spectral resolution FWHM~$\approx
375 \mkms$, a dispersion of $\approx 2$\AA\ per pixel,
 and a nominal wavelength coverage $\lambda \approx
3800$--9000\AA.  Our standard integration was a pair of 1800s
exposures,
and most of the data were obtained under clear conditions.
Longer total exposure times were used
for the faintest fields or in worse conditions.  These exposure
times generally achieved a signal-to-noise S/N~$>3$ per pixel for even
the faintest objects.  
Table~\ref{tab:wfccd} shows a log of the WFCCD observations.

All of these WFCCD data were reduced with a custom IDL software
package\footnote{http://www.ucolick.org/$\sim$xavier/WFCCD/index.html}
\citep[see][for a full description]{pwc+06}.
The extracted and coadded 1D spectra were analyzed with
a modified version of the SDSS pipeline task {\it zfind} to establish
the object's redshift.  This routine fits PCA templates derived from
the SDSS dataset \citep[Early Data Release;][]{stoughton02} 
to the emission and/or absorption lines in the
object spectrum.  In general, these redshift values have
uncertainties of $\sigma_z \approx 10^{-4}$ (i.e.\ $\sigma_v \approx
30\mkms$).
Each of the fits was inspected, and, when necessary, the redshift was 
modified to account for an obvious failure.  
With few exceptions, we determined an unambiguous redshift for every 
object on a given mask.  The exceptions include rare cases
where slit design or fabrication failed (e.g.\ significant portions of
one slit overlapped another) and also faint sources that had no obvious
spectral features.  
This PCA analysis yields eigenvalues for the four galaxy eigenspectra
fitted to our data.  As in \cite{pwc+06}, we define $E_C$ to be the
eigenvalue for the first eigenfunction (which is dominated by
early-type spectral features) and $L_C$ is the the second eigenvalue
minus the sum of the third and fourth coefficients.
All of the spectra are publicly available\footnote{
http://www.ucolick.org/$\sim$xavier/WFCCDOVI/index.html}.

%%%%%%%%%%%%%%%%%%%%%%%%%%%%%%%%%%%%%%%%%%%%5
\begin{figure}[hb]
%\epsscale{0.8}
%\plotone{Figures/fig_stargal.ps}
\includegraphics[height=3.5in,angle=90]{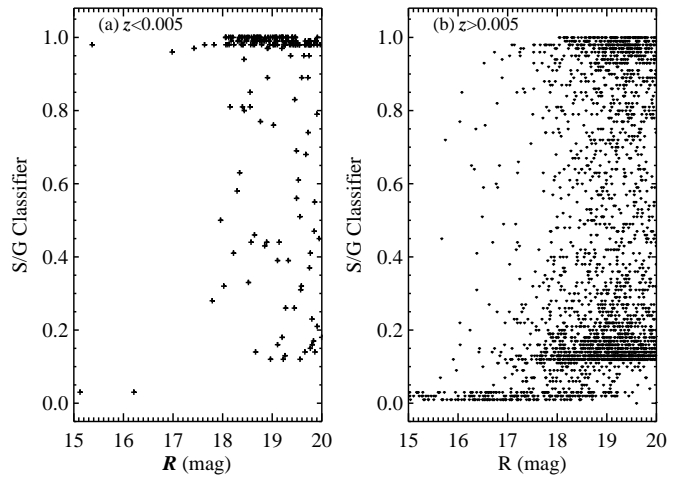}
\caption{SExtractor S/G classifier as a function of the objects apparent $R$-band
  magnitude and having spectroscopic redshifts (a) $z < 0.005$
  (predominantly Galactic stars) and (b) $z>0.005$ (exclusively
  galaxies).
}
\label{fig:stargal}
\end{figure}
%%%%%%%%%%%%%%%%%%%%%%%%%%%%%%%%%%%%%%%%%%%%5

In Figure~\ref{fig:stargal}, we plot the S/G classifier from
SExtractor for the targets with spectroscopic redshifts (a) $z<0.005$,
which are predominantly Galactic stars; and (b) $z>0.005$, which are
exclusively distant galaxies.   As one expects, the majority of
the $z<0.005$ objects have S/G near unity; the exceptions are
primarily objects blended with some other source which was not
de-blended by the SExtractor algorithm.  In contrast, the
extragalactic sources primarily have low S/G classifiers although we
note that $8\%$ have S/G~$>0.98$.  We caution that our
targeting criteria may have missed a subset of very compact galaxies.

\begin{deluxetable*}{rccccrr}
\tablewidth{0pc}
\tablecaption{Summary of Galaxies at $z<z_{\rm em} + 0.01$\label{tab:all_close}}
\tabletypesize{\scriptsize}
\tablehead{\colhead{ID} & \colhead{RA} & \colhead{DEC} & \colhead{$z_{\rm gal}$} & 
\colhead{$L^a$} & \colhead{$\rho^b$} & \colhead{Type$^c$} \\
 & (J2000) & (J2000) & & ($L*$) & 
($\mathrm{h}_{72}^{-1}\,\mathrm{kpc}$)  }
\startdata
\cutinhead{Q0026+1259: $0.005 < z < (z_{\rm em}-0.01)$}
Q0026+1259\_1303 & 00:29:09.2&+13:16:28&0.03295 &  0.02 &    44 &Late\\
Q0026+1259\_1500 & 00:29:01.3&+13:13:12&0.03346 &  0.03 &   158 &Late\\
Q0026+1259\_1143 & 00:29:15.3&+13:20:57&0.03931 &  1.38 &   214 &Early\\
Q0026+1259\_1140 & 00:29:16.4&+13:21:52&0.08043 &  0.11 &   497 &Unkn\\
Q0026+1259\_1722 & 00:28:53.5&+13:24:19&0.05645 &  0.56 &   595 &Late\\
Q0026+1259\_1103 & 00:29:17.7&+13:23:26&0.08073 &  0.08 &   634 &Unkn\\
Q0026+1259\_1901 & 00:28:45.5&+13:20:08&0.08062 &  0.06 &   695 &Early\\
Q0026+1259\_902 & 00:29:23.4&+13:09:41&0.11253 &  1.09 &   781 &Late\\
Q0026+1259\_1125 & 00:29:16.0&+13:10:03&0.13138 &  0.18 &   791 &Unkn\\
Q0026+1259\_512 & 00:29:37.5&+13:10:21&0.09614 &  1.96 &   821 &Early\\
Q0026+1259\_108 & 00:29:53.3&+13:10:50&0.07379 &  0.19 &   879 &Early\\
Q0026+1259\_2045 & 00:28:39.0&+13:19:02&0.10711 &  0.14 &  1007 &Early\\
Q0026+1259\_2141 & 00:28:34.3&+13:07:05&0.10499 &  0.36 &  1437 &Early\\
\cutinhead{Q0026+1259: $(z_{\rm em}-0.01) < z < (z_{\rm em}+0.01)$}
Q0026+1259\_1095 & 00:29:17.9&+13:16:50&0.14736 &  0.20 &   186 &Early\\
Q0026+1259\_1341 & 00:29:07.6&+13:16:06&0.14599 &  0.95 &   219 &Early\\
Q0026+1259\_989 & 00:29:21.2&+13:16:44&0.14802 &  0.27 &   285 &Late\\
Q0026+1259\_1013 & 00:29:20.2&+13:17:15&0.14651 &  2.79 &   286 &Early\\
Q0026+1259\_963 & 00:29:22.2&+13:16:06&0.14183 &  0.21 &   294 &Early\\
Q0026+1259\_1148 & 00:29:15.4&+13:13:25&0.14487 &  0.17 &   381 &Late\\
Q0026+1259\_925 & 00:29:23.4&+13:14:39&0.14480 &  0.29 &   398 &Late\\
Q0026+1259\_764 & 00:29:28.7&+13:14:32&0.14553 &  1.06 &   575 &Late\\
Q0026+1259\_1110 & 00:29:17.2&+13:20:01&0.14655 &  0.49 &   582 &Late\\
Q0026+1259\_1421 & 00:29:04.4&+13:12:17&0.14009 &  0.95 &   614 &Early\\
\enddata
\tablenotetext{a}{\!\,Luminosity relative to L$^*$ as measured from the apparent $R$-band magnitude.  See text for a detailed description.}
\tablenotetext{b}{Impact parameter in physical distance.}
\tablenotetext{c}{The galaxy type is a crude assessment of the spectrum based on a fit of eigenfunctions.  See text for a detailed description.}
\tablecomments{[The complete version of this table is in the electronic edition of the Journal.  The printed edition contains only a sample.]}
\end{deluxetable*}

\section{The Survey Fields}
\label{sec:fields}

In this section, we present the imaging data, photometry, and results
from the WFCCD spectroscopy for each field.  We also comment briefly
on any obvious associations between the galaxies we detect and known (published)
absorbers along the quasar sightlines.
We stress that about half of these fields have been survey for
galaxies by other efforts (e.g.\ SDSS).  Our dicussion, however, is
primarily limited to the galaxies from our LCO/WFCCD survey.
Table~\ref{tab:all_close} summarizes properties of the galaxies discovered in the 20
fields restricted to $0.005 < z < (z_{\rm em} + 0.01$).  
Tables~\ref{tab:Q0026+1259}--\ref{tab:PKS2155-304} 
summarize the photometry and spectroscopy of objects in each field.

In the following, we convert the apparent $R$-band magnitude of each
galaxy with a measured redshift into a luminosity relative to $L^*$.
For the latter, we adopt the Sloan $r$-band value from \cite{blanton03}
as measured from the Sloan Digital Sky Survey which corresponds to $z
\approx 0.1$: $r^* = -20.44 + 5\log h_{100}$.  With the cosmology we
have adopted, this yields $r^* = -21.12$\,mag.  

For each galaxy, we assess the spectral type from the $E_C$ and $L_C$
coefficients as follows (based in part on a visual inspection of several tens
of the galaxies): 

\begin{itemize}
\item Early-type:  $E_C > 0.6$ and $L_C < 0.6$
\item Late-type:   $E_C < 0.8$ and $L_C > 0.4$, or $E_C<0$ and $L_C >
  0.2$, or $E_C < 0$ and $L_C < 0$. 
\end{itemize}
Galaxies with any other pair of coefficient values are labelled
`unknown'.
For every galaxy with an impact parameter $\rho < 300$\,kpc we have
visually verified each spectral type, placing each
into either the late or early-type category.

We then imposed a $k$-correction to the apparent
magnitude using a redshifted template spectrum integrated with an
$R$-band filter.  We use an E/S0 template for the early-type galaxies,
and an Sc template for the late-type systems.  No correction was made
for galaxies classified as unknown.  This $k$-correction
generally amounts to only a tenth or few tenths correction to the
magnitudes, with larger values for more distant galaxies $(z>0.2)$.
Finally, we converted the $k$-corrected, apparent $R$-band magnitudes
into luminosities relative to $L^*$ with the luminosity distance
calculated using the assumed WMAP cosmology.

In total, we obtained spectra for \nspec\ sources and confirmed
\nztot\ redshifts.  Of these, there are \nzigm\  galaxies with
$0.005 < z < (\mzem - 0.01)$ which are suitable for studying the
association between galaxies and absorption lines in the low-$z$ IGM.
The properties of these galaxies are summarized in 
Table~\ref{tab:all_close}. 
Lastly, Table~\ref{tab:fields} lists the spectroscopic completeness of
each field as a function of magnitude limit and angular offset from
the quasar, e.g.\ $C_{19.5}^{5'}$ gives the completeness percentile to
$R=19.5$\,mag and $\theta_{\rm max} = 5'$ offset from the quasar.

%%%%%%%%%%%%%%%%%%%%%%%%%%%%%%%
\begin{figure}
%\epsscale{0.8}
%\plotone{Figures/Q0026+1259_img.ps}
\includegraphics[width=3.5in]{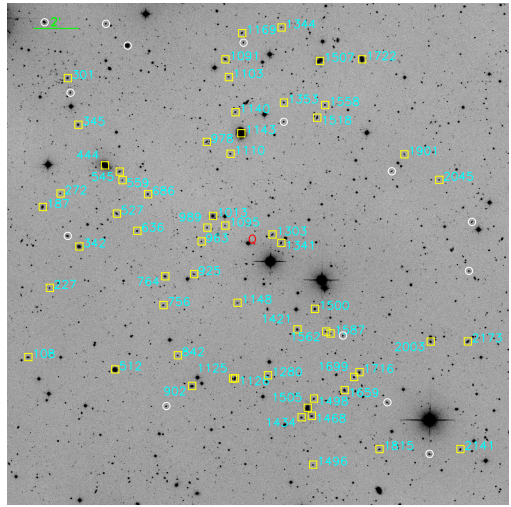}
\caption{$R$-band image of the field surrounding
Q0026+1259 ($z_{\rm em} = 0.142$; marked with a red ``Q'').  
All galaxies with $z<(z_{\rm em}+0.01)$ are labelled with their
redshifts.
Unless otherwise indicated, labels are placed to the right
and just above the galaxy.  Objects circled in pink 
were targeted (i.e.\ $R<19.5$\,mag and the imaging indicated it was
not a point source)
but no redshift was determined either because no mask included
it or a secure redshift could not be determined from the spectrum.
}
\label{fig:Q0026+1259_img}
\end{figure}

\begin{figure}
%\epsscale{0.8}
\includegraphics[height=3.5in,angle=90]{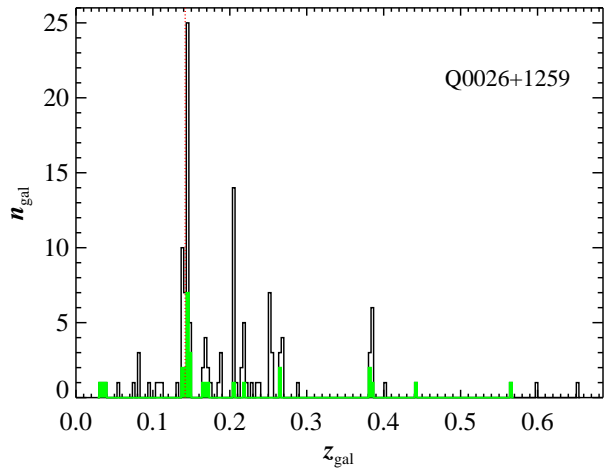}
%%%
\caption{Histogram of galaxy redshifts for the field surrounding
  Q0026+1259 ($z_{\rm em} = 0.142$), binned in intervals of $\Delta z
  = 0.00333$.  The solid (green) histogram shows the distribution for
  galaxies within $5'$ of the quasar while the open (black) histogram
  is for all galaxies surveyed in the field.
  The dotted (red) vertical line indicates the quasar emission-line redshift
  ($z_{\rm em}$). 
}
\label{fig:Q0026+1259_hist}
\end{figure}

\begin{deluxetable*}{rrrcccccc}
\tablewidth{0pc}
\tablecaption{Q0026+1259: Object Summary \label{tab:Q0026+1259}}
\tabletypesize{\footnotesize}
\tablehead{\colhead{ID} & \colhead{RA} & \colhead{DEC} 
& \colhead{$B$}
& \colhead{$R$} & \colhead{S/G$^a$} 
& \colhead{Area}& \colhead{flg$^b$} & \colhead{$z$} \\
 & & & (mag) & (mag) &  & ($\square''$) }
\startdata
1&00:28:40.3&+13:08:22&$11.51 \pm  0.11$&$10.20 \pm  0.08$&0.69&  76.8& 0& \nodata \\
2&00:28:27.4&+13:16:27&$18.79 \pm  0.12$&$17.43 \pm  0.08$&0.96&   4.8& 0& \nodata \\
3&00:28:27.6&+13:26:27&$19.12 \pm  0.12$&$17.33 \pm  0.08$&1.00&   2.6& 0& \nodata \\
4&00:28:26.8&+13:15:39&$19.54 \pm  0.12$&$18.21 \pm  0.08$&1.00&   2.4& 0& \nodata \\
5&00:28:27.2&+13:24:49&$20.46 \pm  0.14$&$18.22 \pm  0.08$&1.00&   2.4& 0& \nodata \\
6&00:28:26.4&+13:14:06&$21.46 \pm  0.23$&$19.60 \pm  0.09$&0.14&   3.6& 0& \nodata \\
7&00:28:26.1&+13:13:18&$21.76 \pm  0.27$&$19.50 \pm  0.09$&1.00&   2.9& 0& \nodata \\
8&00:28:25.8&+13:10:15&$23.74 \pm  1.00$&$21.42 \pm  0.15$&0.19&   2.4& 0& \nodata \\
9&00:28:25.8&+13:10:33&$21.64 \pm  0.24$&$19.64 \pm  0.09$&0.87&   2.6& 0& \nodata \\
10&00:28:25.9&+13:11:33&$20.69 \pm  0.16$&$18.26 \pm  0.08$&1.00&   2.4& 0& \nodata \\
\enddata
\tablenotetext{a}{Star/galaxy classifier calculated by SExtractor.
Values near unity indicate a stellar-like point-spread function.}
\tablenotetext{b}{This binary flag has the following code:
1: Survey target; 2: Spectrum taken; 4: Redshift measured.}
\tablecomments{[The complete version of this table is in the electronic edition of the Journal.  The printed edition contains only a sample.]}
\end{deluxetable*}

\subsection{Q0026+1259}

To date this quasar has no published list of IGM absorption lines, 
presumably because its UV spectroscopic dataset is relatively sparse.  
%As
%described below, we have analyzed the 20\,ks of {\it FUSE} data for
%possible \ion{O}{6} absorption at selected galaxy redshifts.
%
Figure~\ref{fig:Q0026+1259_img} shows the field and marks the
galaxies surrounding the quasar with measured redshifts.
Figure~\ref{fig:Q0026+1259_hist} shows a histogram of the 
redshifts of the galaxies in the field surrounding the quasar.
There is a relatively large `spike' at the quasar
redshift ($z_{\rm em} = 0.142$) which marks a galaxy group (several $L
\approx L*$ galaxies lie within $\rho=300$\,kpc).  One
identifies no other significant large-scale structures foreground to
the quasar in our LCO/WFCCD survey.
Table~\ref{tab:Q0026+1259} summarizes the photometry and spectroscopy
for galaxies and stars in the field surrounding Q0026+1259. 

%%%%%%%%%%%%%%%%%%%%%%%%%%%%
\subsection{\tonso}

The \tonso\ sightline has been surveyed for \lya\ absorption by
\cite[][STIS/G140M spectra]{pss04} and for \ion{O}{6} absorption by
\cite[][{\it FUSE}]{dsr+06}.  The latter report \ion{O}{6}
absorbers at $z=0.0234, 0.0436$ and $z=0.0456$, each of which shows a
galaxy within an impact parameter of $\rho < 300$\,kpc and within a
velocity offset of $|\delta v| < 300 \mkms$.  None of these were reported in
\cite{stockeetal06}\footnote{Indeed, they remarked that the \ovi\ absorber
  at $z=0.436$ was the ``most isolated'' of their sample with no
  bright galaxy within 1\,Mpc.} 
presumably because these galaxies were fainter than the magnitude
limit of their galaxy compilation.
\cite{wakker09} also associated galaxies at $z\approx 0$ with \lya\
and \ovi\ absorption along this sightline.

Figure~\ref{fig:TONS180_img} shows the field and marks the
galaxies surrounding the quasar with measured redshifts.
Given the low emission redshift of the quasar, there are only a
handful of galaxies identified in its foreground.  One also notes
several bright galaxies coincident with the quasar redshift and at
relatively small impact parameter ($\rho \approx 300$\,kpc).
Figure~\ref{fig:TONS180_hist} shows a histogram of the 
redshifts of the galaxies in the field surrounding the quasar.
Table~\ref{tab:TonS180} summarizes the photometry and spectroscopy
of our survey for galaxies and stars in the field surrounding \tonso.

%%%%%%%%%%%%%%%%%%%%%%%%%%%%
\subsection{\tonst}

Although the quasar \tonst\ has an impressive UV spectroscopic
dataset, there is no published 
survey for IGM absorption beyond the very local
universe \citep{wakker09}.  Coincidentally, we have discovered very few
galaxies foreground to the quasar to our magnitude limit
(Figure~\ref{fig:TONS210_img}, Table~\ref{tab:all_close}) despite
obtaining spectra for 71 extragalactic sources
(Figure~\ref{fig:TONS210_hist}, Table~\ref{tab:TonS210}).  None of the
foreground sources lie at close impact parameter. %and we discuss this
%field no further in this paper.

%%%%%%%%%%%%%%%%%%%%%%%%%%%%
\subsection{PKS0312$-$770}

We previously published the spectrum of a galaxy associated with the
$z \approx 0.21$ Lyman limit system toward PKS0312$-$770
\citep{lpk+09}.   The sightline has also been surveyed by several
groups for \lya\ and \ovi\ absorption
\citep{ds08,tc08a,tripp08}. \cite{tripp08} report two \ovi\
absorbers at $z=0.1589$ and $z=0.19827$ where we identify bright
galaxies in our survey but only at large impact parameters ($\rho > 500$\,kpc).    
We also identify a galaxy at $z=0.0736$ with $\rho =
954$\,kpc that may be associated with the `void' absorber reported by
\cite{stockeetal06}.  It has a luminosity $L \approx 0.1L*$ which places
it below their survey limit.

Tables~\ref{tab:PKS0312-770} 
summarizes the photometry and spectroscopy
for galaxies and stars in the field surrounding PKS0312--770.
Figure~\ref{fig:PKS0312-770_img} displays our imaging data which
covers a rectangular $15' \times 32'$ field of view, and
Figure~\ref{fig:PKS0312-770_hist} shows a redshift histogram 
for the field. 

%%%%%%%%%%%%%%%%%%%%%%%%%%%%
\subsection{PKS0405--123}

This field has received significant attention over the past decade,
largely because of its relatively high redshift but also because it
exhibits several strong \ovi\ absorbers
\citep{cp00,pks0405_uv,cpw+05,pwc+06,willigeretal06,tripp08,howk+09,wakker09,snw+10}.  
We published the results of our galaxy survey in \cite{cpw+05} and
\cite{pwc+06}; here we
only summarize properties of the galaxies close to the sightline 
(Table~\ref{tab:all_close}).  We also remind the reader that our
galaxy survey had non-uniform spectral and imaging coverage beyond the
inner $\approx 20' \times 20'$ field-of-view.

%%%%%%%%%%%%%%%%%%%%%%%%%%%%
\subsection{PKS0558--504}

This quasar has a very limited UV spectral dataset, and the only
IGM analysis has been performed at $z \approx 0$ \citep{wakker09}.
By chance, our survey also shows no galaxies 
at low impact parameters (Tables~\ref{tab:PKS0558-504} and 
Table~\ref{tab:all_close}).
Figure~\ref{fig:PKS0558-504_img} shows the field and marks the
galaxies surrounding the quasar with measured redshifts and 
Figure~\ref{fig:PKS0558-504_hist} presents a redshift histogram. 

%%%%%%%%%%%%%%%%%%%%%%%%%%%%
\subsection{PG1004+130}

This sightline has been analyzed for \lya\ absorption by
\cite{bowen+02} who associated a cluster of low $z$ \lya\ absorption lines
with a bright nearby galaxy (UGC~5454).  A relatively high quality 
{\it FUSE} dataset exists but has not yet been surveyed for \ovi\
absorption. 
As discussed in Prochaska et al. (in prep), we have also analyzed these UV
datasets for \lya\ and \ovi\ absorption at the redshifts of the
galaxies in our LCO/WFCCD survey
with small impact parameter to the sightline 
(Table~\ref{tab:all_close}).    %In particular there are three
%galaxies with $z \approx 0.009$ and $\rho < 100$\,kpc that are
%associated with a `cluster' of \lya\ absorption in the STIS/G140M spectra.

The photometry and spectroscopy of
our full survey is listed in 
Table~\ref{tab:PG1004+130} and
Figure~\ref{fig:PG1004+130_img} shows the field and marks the
galaxies surrounding the quasars with determined redshifts.
Figure~\ref{fig:PG1004+130_hist} provides a redshift histogram.

%%%%%%%%%%%%%%%%%%%%%%%%%%%%
\subsection{HE1029-140}

This quasar has a modest UV spectral dataset and has been analyzed for
\lya\ absorption by \cite{pss04}.  
Figure~\ref{fig:HE1029-140_img} shows the field and marks the
few galaxies with $z<z_{\rm em}$ that we have identified in the
LCO/WFCCD survey.
Table~\ref{tab:HE1029-140} summarizes the photometry and spectroscopy
of all objects in the field and 
Figure~\ref{fig:HE1029-140_hist} gives a redshift histogram of 
the galaxies from our survey.

%%%%%%%%%%%%%%%%%%%%%%%%%%%%
\subsection{PG1116+215}

This sightline has been studied by \cite{tripp+98} and \cite{sts+04}  
for absorption lines and associated galaxies, along with several other
more recent analyses \citep{tripp08,ds08,wakker09}. 
They identify three \ovi\ systems foreground to the quasar at $z=0.0597,
0.13879$ and 0.1655.  Each of these is associated with a galaxy at
$\rho \le 150$\,kpc in our survey.  For the $z=0.0597$ system, we
identify 6 galaxies within 1\,Mpc including several with $L \gtrsim
L^*$ which suggests a group environment (Table~\ref{tab:all_close}).
None of these galaxies were identified in previous surveys.

Table~\ref{tab:PG1116+215} summarizes the photometry and spectroscopy
for the objects that we surveyed in the field surrounding PG1116+215.
Figure~\ref{fig:PG1116+215_img} presents an image of the field and marks the
galaxies surrounding the quasar with determined $z<z_{\rm em}$ and
Figure~\ref{fig:PG1116+215_hist} shows a redshift histogram 
of the full galaxy survey.

%%%%%%%%%%%%%%%%%%%%%%%%%%%%
\subsection{PG1211+143}

\cite{tsg+05} analyzed this field for IGM absorption lines and
corresponding galaxies.  They reported a pair of \ovi\ systems at
$z=0.0513$ and 0.0645 and associated these with $L \approx L*$ galaxies
at $\rho \approx 150$\,kpc.   Our survey includes these two galaxies
(Table~\ref{tab:all_close}) and an additional, fainter galaxy
at significantly smaller impact parameter for the system at $z =
0.0646$ ($\rho = 70$\,kpc).  
We also identify additional galaxies associated with the galaxy group
at $z=0.0513$. 
\cite{wakker09} also analyzed this field, focusing on $z \approx 0$.

Table~\ref{tab:PG1211+143} summarizes the photometry and spectroscopy
for galaxies and stars in the field surrounding PG1211+143.
Figure~\ref{fig:PG1211+143_img} shows the field and marks the
galaxies surrounding the quasar while
Figure~\ref{fig:PG1211+143_hist} gives a galaxy redshift histogram. 

%%%%%%%%%%%%%%%%%%%%%%%%%%%%
\subsection{PG1216+069}

This sightline has been studied for absorption by multiple groups
\citep{tripp08,tc08a,ds08,cm09,wakker09} with \ovi\ systems reported at
$z=0.1236, 0.268$ and 0.282.  Our survey reveals galaxies at every
redshift, but only the lowest redshift system shows an example with
small impact parameter ($\rho = 85$\,kpc).  In addition, we report an
overdensity of galaxies at $z=0.0805$ (5 galaxies within 1\,Mpc
including an $L^*$ galaxy; Table~\ref{tab:all_close}) which is
associated with strong \lya\ absorption but no apparent \ovi\ gas to a
sensitive limit.

The sightline is also notable for intersecting Virgo, and
\cite{tripp+05} report on a strong \ion{H}{1} absorber near the
NGC~4261 group ($z=0.0063$).  No apparent \ovi\ absorption is detected
at the redshift
of either structure although the equivalent width limits are poor from the {\it
  FUSE} spectra at these wavelengths.

Table~\ref{tab:PG1216+069} summarizes the photometry and spectroscopy,
Figure~\ref{fig:PG1216+069_img} shows the field and marks the
galaxies surrounding the quasars with determined redshifts.
Figure~\ref{fig:PG1216+069_hist} shows a histogram of the 
redshifts of the galaxies in the field surrounding the quasar.

%%%%%%%%%%%%%%%%%%%%%%%%%%%%
\subsection{3C273}

Because of its very high UV flux, this quasar and the field around it
have been the subject of many previous studies including some of the first
galaxy surveys along quasar sightlines \citep{mwd+93}.
\cite{tripp08} report \ovi\ detections at several redshifts, all at
column densities that lie below the sensitivity limit of most other UV
spectral datasets, i.e.\ $N({\rm O^{+5}}) < 10^{13.5} \cm{-2}$.  None of
these absorbers shows a corresponding galaxy in our survey with $\rho
< 300$\,kpc, but we do identify galaxies at larger impact parameter
(Table~\ref{tab:all_close}).  
\cite{wakker09} also analyzed this field, focusing on $z \approx 0$.

Table~\ref{tab:3C273} summarizes the photometry and spectroscopy
for galaxies and stars in the field surrounding 3C273.
Figure~\ref{fig:3C273_img} shows the field and marks the
galaxies surrounding the quasars with determined redshifts.
Figure~\ref{fig:3C273_hist} shows a histogram of the 
redshifts of the galaxies in the field surrounding the quasar.

%%%%%%%%%%%%%%%%%%%%%%%%%%%%
\subsection{Q1230+095}

This quasar has a very sparse UV spectral dataset and there is no
published list of IGM absorption.  We emphasize that the quasar's redshift
($z_{\rm em} = 0.415$) make it a very promising sightline for studying
the IGM/galaxy connection.  Furthermore, there is a set of galaxies
at small impact parameters revealed by our survey
(Figure~\ref{fig:Q1230+095_img}, Table~\ref{tab:all_close}).  We
encourage observations with the {\it HST}/COS instrument.
Table~\ref{tab:Q1230+095} summarizes the full photometry and spectroscopy
for objects in the field surrounding Q1230+095 and 
Figure~\ref{fig:Q1230+095_hist} histograms the 
galaxy redshifts.

%%%%%%%%%%%%%%%%%%%%%%%%%%%%
\subsection{PKS1302-102}

This field has been surveyed for \lya\ and \ovi\
absorption by several groups
\citep{dsr+06,cpc+08,ds08,tc08a,tripp08,wakker09}.  
These groups have reported \ovi\ detections at multiple redshifts:
$z=0.0423, 0.0647, 0.0940, 0.0989, 0.1453, 0.1916, 0.2256$, and 0.2274. 
Our survey has revealed nearly 100 galaxies foreground to the quasar
including several at impact parameters $\rho < 100$\,kpc
\citep[see also][]{cpc+08}.
We detect galaxy within 300\,kpc and with $|\delta v| < 400 \mkms$
for each absorber at $z<0.2$ except at $z=0.0989$.  This includes two sub-$L*$
galaxies that lie at low impact parameter $(\rho < 100$\,kpc;
Table~\ref{tab:all_close}, Figure~\ref{fig:PKS1302-102_img}).
Table~\ref{tab:PKS1302-102} summarizes the photometry and spectroscopy
for galaxies and stars in the field and 
Figure~\ref{fig:PKS1302-102_hist} presents a redshift histogram of all 
galaxies. 

%%%%%%%%%%%%%%%%%%%%%%%%%%%%
\subsection{PG1307+085}

This quasar has only a sparse dataset of UV spectroscopy and there is
no published list for \lya\ or \ovi\ absorption systems.  Its relatively
high redshift ($z_{\rm em} = 0.155$) marks it as a valuable sightline for future study and
our survey provides a modest set of galaxies for cross-correlation
analysis (Table~\ref{tab:all_close}). 
The full photometry and spectroscopy for objects from our survey in the field are
listed in Table~\ref{tab:PG1307+085}. 
Figure~\ref{fig:PG1307+085_img} presents an $R$-band image
of the field and marks the surrounding
galaxies with $z< z_{\rm em}$. 
Figure~\ref{fig:PG1307+085_hist} shows a redshift histogram for the 
galaxies we have discovered.  

%%%%%%%%%%%%%%%%%%%%%%%%%%%%
\subsection{MRK1383}

Despite its low redshift and correspondingly small pathlength,
this sightline has been studied extensively for \lya\ and \ovi\
absorption \citep{dsr+06,ds08,wakker09}.  
Our survey reveals only a handful of galaxies with $z < z_{\rm em}$
(Table~\ref{tab:all_close}, Figure~\ref{fig:MRK1383_img}), but we also note
that this field has one of the lowest completeness levels to
$R=19.5$\,mag (80\%). 
Table~\ref{tab:MRK1383} summarizes the photometry and spectroscopy
for galaxies and stars in the field surrounding MRK1383.
Figure~\ref{fig:MRK1383_img} provides an image of the field and marks the
foreground galaxies, and 
Figure~\ref{fig:MRK1383_hist} shows a redshift histogram for
the galaxies. 

%%%%%%%%%%%%%%%%%%%%%%%%%%%%
\subsection{Q1553+113}

This quasar boasts only a short {\it FUSE} exposure and no additional
UV spectroscopy.  It has been surveyed for IGM absorption only at $z \approx 0$ by
\cite{wakker09}.  
Given the extensive dataset of galaxies that we have
discovered in the foreground  (Table~\ref{tab:all_close},
Figure~\ref{fig:Q1553+113_img}), we encourage future
observations with {\it HST}/COS. 
The photometry and spectroscopy for objects from our survey in the field are
listed in Table~\ref{tab:Q1553+113} and
a redshift histogram for the galaxies is provided by
Figure~\ref{fig:Q1553+113_hist}.

%%%%%%%%%%%%%%%%%%%%%%%%%%%%
\subsection{PKS2005-489}

This sightline has been surveyed for \lya\ and \ovi\ absorption
over the relatively short pathlength provided by this $z=0.071$ AGN
\citep{pss04,dsr+06,ds08}.   
There are no \ovi\ systems reported.
Our survey reveals two sets of
foreground galaxies at redshifts $z \approx 0.045$ and 0.057, but none at very
close impact parameter to the sightline
(Figure~\ref{fig:PKS2005-489_img}, Table~\ref{tab:all_close}).
Table~\ref{tab:PKS2005-489} summarizes the photometry and spectroscopy
for all objects in the field surrounding PKS2005--489.
Figure~\ref{fig:PKS2005-489_hist} presents a redshift histogram for
all of the galaxies in our survey.

%%%%%%%%%%%%%%%%%%%%%%%%%%%%
\subsection{FJ2155-092}

The quasar FJ2155--092, also known as PHL~1811, has a high quality UV
spectral dataset.  \cite{jbt05} has published an analysis of the
Lyman limit system at $z=0.081$ and its neighboring
galaxies.  Other groups have examined the spectra for \lya\ and \ovi\
absorption \citep{tripp08,ds08,wakker09}. 
Altogether, the sightline boasts four \ovi\ systems at $z=0.0788,
0.1326, 0.1581$ and 0.1769, all of which have a neighboring galaxy in
our survey with
$\rho < 350$\,kpc and $|\delta v| < 400 \mkms$ but none at
very low impact parameter.
Figure~\ref{fig:FJ2155-092_img} shows the field and marks the
galaxies foreground to the quasar with determined redshifts.
Table~\ref{tab:FJ2155-092} lists the photometry and spectroscopy
for galaxies and stars throughout the field and 
a redshift histogram of the 
galaxies is given by Figure~\ref{fig:FJ2155-092_hist}.

%%%%%%%%%%%%%%%%%%%%%%%%%%%%
\subsection{PKS2155-304}

This sightline has received significant attention, primarily because
of the claimed (and debated) \ion{O}{7} and \ion{O}{8} absorption 
at $z \approx 0.055$ \citep{fangetal2002,cagnoni+03}.  The high quality UV
spectral datasets have also been surveyed for \lya\ and \ovi\
absorption 
\citep[who report detections at $z=0.0541$ and
0.0571][]{dsr+06,ds08,wakker09}. 
We associate each of these with a bright ($L>L*$)
galaxy in our survey but at relatively large impact parameter ($\rho \sim 500$\,kpc).
The photometry and spectroscopy for our survey of the
the field is given in Table~\ref{tab:PKS2155-304}.
We show an image of the field and mark the foreground galaxies in
Figure~\ref{fig:PKS2155-304_img} and present a redshift histogram
for all of the galaxies in Figure~\ref{fig:PKS2155-304_hist}. 
We note that our deeper survey reveals no additional galaxies at
$z \approx 0.055$ from the ones first reported by
\cite[see Table~\ref{tab:all_close};][]{shulletal03}. 

%%%
% Q0026+1259

%%%
% Ton S 180
\begin{figure}
%\epsscale{0.8}
%\plotone{Figures/TONS180_img.ps}
%\includegraphics[width=6in]{f4.eps}
\caption{Same as for Figure~\ref{fig:Q0026+1259_img} but for the
  field surrounding \tonso\ ($z_{\rm em} = 0.062$).
}
\label{fig:TONS180_img}
\end{figure}

\begin{figure}
%\epsscale{0.8}
%\plotone{Figures/TONS180_hist.ps}
%\includegraphics[width=6in]{f5.eps}
\caption{Same as for Figure~\ref{fig:Q0026+1259_hist} but for the
  field surrounding \tonso\ ($z_{\rm em} = 0.062$).
}
\label{fig:TONS180_hist}
\end{figure}

%%%
% Ton S 210
\begin{figure}
%\epsscale{0.8}
%\plotone{Figures/TONS210_img.ps}
%\includegraphics[width=6in]{f6.eps}
\caption{Same as for Figure~\ref{fig:Q0026+1259_img} but for the
  field surrounding \tonst\ ($z_{\rm em} = 0.116$).
}
\label{fig:TONS210_img}
\end{figure}

\begin{figure}
%\epsscale{0.8}
%\plotone{Figures/TONS210_hist.ps}
%\includegraphics[width=6in]{f7.eps}
\caption{Same as for Figure~\ref{fig:Q0026+1259_hist} but for the
  field surrounding \tonst\ ($z_{\rm em} = 0.116$).
}
\label{fig:TONS210_hist}
\end{figure}

%%%%%%%%%%%%%%%%%%%
% PKS0312-770
\begin{figure}
%\epsscale{0.8}
%\plotone{Figures/PKS0312-770_img.ps}
%\includegraphics[width=6in]{f8.eps}
\caption{Same as for Figure~\ref{fig:Q0026+1259_img} but for the
  field surrounding PKS0312--770 ($z_{\rm em} = 0.223$).
}
\label{fig:PKS0312-770_img}
\end{figure}

\begin{figure}
%\epsscale{0.8}
%\plotone{Figures/PKS0312-770_hist.ps}
%\includegraphics[width=6in]{f9.eps}
\caption{Same as for Figure~\ref{fig:Q0026+1259_hist} but for the
  field surrounding PKS0312--770 ($z_{\rm em} = 0.223$).
}
\label{fig:PKS0312-770_hist}
\end{figure}

%%%
% PKS0558-504
\begin{figure}
%\epsscale{0.8}
%\plotone{Figures/PKS0558-504_img.ps}
%\includegraphics[width=6in]{f10.eps}
\caption{Same as for Figure~\ref{fig:Q0026+1259_img} but for the
  field surrounding PKS0558--504 ($z_{\rm em} = 0.137$).
}
\label{fig:PKS0558-504_img}
\end{figure}

\begin{figure}
%\epsscale{0.8}
%\plotone{Figures/PKS0558-504_hist.ps}
%\includegraphics[width=6in]{f11.eps}
\caption{Same as for Figure~\ref{fig:Q0026+1259_hist} but for the
  field surrounding PKS0558--504 ($z_{\rm em} = 0.137$).
}
\label{fig:PKS0558-504_hist}
\end{figure}

%%%%%%%%%%%%%%%%%%%%%%%
% PG1004+130
\begin{figure}
%\epsscale{0.8}
%\plotone{Figures/PG1004+130_img.ps}
%\includegraphics[width=6in]{f12.eps}
\caption{Same as for Figure~\ref{fig:Q0026+1259_img} but for the
  field surrounding PG1004+130 ($z_{\rm em} = 0.240$).
}
\label{fig:PG1004+130_img}
\end{figure}

\begin{figure}
%\epsscale{0.8}
%\plotone{Figures/PG1004+130_hist.ps}
%\includegraphics[width=6in]{f13.eps}
\caption{Same as for Figure~\ref{fig:Q0026+1259_hist} but for the
  field surrounding PG1004+130 ($z_{\rm em} = 0.240$).
}
\label{fig:PG1004+130_hist}
\end{figure}

%%%%%%%%%%%%%%%%%%%%%%%
% HE1029-140
\begin{figure}
%\epsscale{0.8}
%\plotone{Figures/HE1029-140_img.ps}
%\includegraphics[width=6in]{f14.eps}
\caption{Same as for Figure~\ref{fig:Q0026+1259_img} but for the
  field surrounding HE1029--140 ($z_{\rm em} = 0.086$).
}
\label{fig:HE1029-140_img}
\end{figure}

\begin{figure}
%\epsscale{0.8}
%\plotone{Figures/HE1029-140_hist.ps}
%\includegraphics[width=6in]{f15.eps}
\caption{Same as for Figure~\ref{fig:Q0026+1259_hist} but for the
  field surrounding HE1029--140 ($z_{\rm em} = 0.086$).
}
\label{fig:HE1029-140_hist}
\end{figure}

%%%%%%%%%%%%%%%%%%%%%%%
% PG1116+215
\begin{figure}
%\epsscale{0.8}
%\plotone{Figures/PG1116+215_img.ps}
%\includegraphics[width=6in]{f16.eps}
\caption{Same as for Figure~\ref{fig:Q0026+1259_img} but for the
  field surrounding PG1116+215 ($z_{\rm em} = 0.177$).
}
\label{fig:PG1116+215_img}
\end{figure}

\begin{figure}
%\epsscale{0.8}
%\plotone{Figures/PG1116+215_hist.ps}
%\includegraphics[width=6in]{f17.eps}
\caption{Same as for Figure~\ref{fig:Q0026+1259_hist} but for the
  field surrounding PG1116+215 ($z_{\rm em} = 0.177$).
}
\label{fig:PG1116+215_hist}
\end{figure}

%%%%%%%%%%%%%%%%%%%%%%%
% PG1211+143
\begin{figure}
%\epsscale{0.8}
%\plotone{Figures/PG1211+143_img.ps}
%\includegraphics[width=6in]{f18.eps}
\caption{Same as for Figure~\ref{fig:Q0026+1259_img} but for the
  field surrounding PG1211+143 ($z_{\rm em} = 0.809$).
}
\label{fig:PG1211+143_img}
\end{figure}

\begin{figure}
%\epsscale{0.8}
%\plotone{Figures/PG1211+143_hist.ps}
%\includegraphics[width=6in]{f19.eps}
\caption{Same as for Figure~\ref{fig:Q0026+1259_hist} but for the
  field surrounding PG1211+143 ($z_{\rm em} = 0.809$).
}
\label{fig:PG1211+143_hist}
\end{figure}

%%%%%%%%%%%%%%%%%%%%%%%
% PG1216+069
\begin{figure}
%\epsscale{0.8}
%\plotone{Figures/PG1216+069_img.ps}
%\includegraphics[width=6in]{f20.eps}
\caption{Same as for Figure~\ref{fig:Q0026+1259_img} but for the
  field surrounding PG1216+069 ($z_{\rm em} = 0.331$).
}
\label{fig:PG1216+069_img}
\end{figure}

\begin{figure}
%\epsscale{0.8}
%\plotone{Figures/PG1216+069_hist.ps}
%\includegraphics[width=6in]{f21.eps}
\caption{Same as for Figure~\ref{fig:Q0026+1259_hist} but for the
  field surrounding PG1216+069 ($z_{\rm em} = 0.331$).
}
\label{fig:PG1216+069_hist}
\end{figure}

\clearpage

%%%%%%%%%%%%%%%%%%%%%%%
% 3C273
\begin{figure}
%\epsscale{0.8}
%\plotone{Figures/3C273_img.ps}
%\includegraphics[width=6in]{f22.eps}
\caption{Same as for Figure~\ref{fig:Q0026+1259_img} but for the
  field surrounding 3C273 ($z_{\rm em} = 0.158$).
}
\label{fig:3C273_img}
\end{figure}

\begin{figure}
%\epsscale{0.8}
%\plotone{Figures/3C273_hist.ps}
%\includegraphics[width=6in]{f23.eps}
\caption{Same as for Figure~\ref{fig:Q0026+1259_hist} but for the
  field surrounding 3C273 ($z_{\rm em} = 0.158$).
}
\label{fig:3C273_hist}
\end{figure}

%%%%%%%%%%%%%%%%%%%%%%%
% Q1230+095
\begin{figure}
%\epsscale{0.8}
%\plotone{Figures/Q1230+095_img.ps}
%\includegraphics[width=6in]{f24.eps}
\caption{Same as for Figure~\ref{fig:Q0026+1259_img} but for the
  field surrounding Q1230+095 ($z_{\rm em} = 0.415$).
}
\label{fig:Q1230+095_img}
\end{figure}

\begin{figure}
%\epsscale{0.8}
%\plotone{Figures/Q1230+095_hist.ps}
%\includegraphics[width=6in]{f25.eps}
\caption{Same as for Figure~\ref{fig:Q0026+1259_hist} but for the
  field surrounding Q1230+095 ($z_{\rm em} = 0.415$).
}
\label{fig:Q1230+095_hist}
\end{figure}

%%%%%%%%%%%%%%%%%%%%%%%
% PKS1302-102
\begin{figure}
%\epsscale{0.8}
%\plotone{Figures/PKS1302-102_img.ps}
%\includegraphics[width=6in]{f26.eps}
\caption{Same as for Figure~\ref{fig:Q0026+1259_img} but for the
  field surrounding PKS1302--102 ($z_{\rm em} = 0.286$).
}
\label{fig:PKS1302-102_img}
\end{figure}

\begin{figure}
%\epsscale{0.8}
%\plotone{Figures/PKS1302-102_hist.ps}
%\includegraphics[width=6in]{f27.eps}
\caption{Same as for Figure~\ref{fig:Q0026+1259_hist} but for the
  field surrounding PKS1302--102 ($z_{\rm em} = 0.286$).
}
\label{fig:PKS1302-102_hist}
\end{figure}

%%%%%%%%%%%%%%%%%%%%%%%
% PG1307+085
\begin{figure}
%\epsscale{0.8}
%\plotone{Figures/PG1307+085_img.ps}
%\includegraphics[width=6in]{f28.eps}
\caption{Same as for Figure~\ref{fig:Q0026+1259_img} but for the
  field surrounding PG1307+085 ($z_{\rm em} = 0.155$).
}
\label{fig:PG1307+085_img}
\end{figure}

\begin{figure}
%\epsscale{0.8}
%\plotone{Figures/PG1307+085_hist.ps}
%\includegraphics[width=6in]{f29.eps}
\caption{Same as for Figure~\ref{fig:Q0026+1259_hist} but for the
  field surrounding PG1307+085 ($z_{\rm em} = 0.155$).
}
\label{fig:PG1307+085_hist}
\end{figure}

%%%%%%%%%%%%%%%%%%%%%%%
% MRK1383
\begin{figure}
%\epsscale{0.8}
%\plotone{Figures/MRK1383_img.ps}
%\includegraphics[width=6in]{f30.eps}
\caption{Same as for Figure~\ref{fig:Q0026+1259_img} but for the
  field surrounding MRK1383 ($z_{\rm em} = 0.086$).
}
\label{fig:MRK1383_img}
\end{figure}

\begin{figure}
%\epsscale{0.8}
%\plotone{Figures/MRK1383_hist.ps}
%\includegraphics[width=6in]{f31.eps}
\caption{Same as for Figure~\ref{fig:Q0026+1259_hist} but for the
  field surrounding MRK1383 ($z_{\rm em} = 0.086$).
}
\label{fig:MRK1383_hist}
\end{figure}

%%%%%%%%%%%%%%%%%%%%%%%
% Q1553+113
\begin{figure}
%\epsscale{0.8}
%\plotone{Figures/Q1553+113_img.ps}
%\includegraphics[width=6in]{f32.eps}
\caption{Same as for Figure~\ref{fig:Q0026+1259_img} but for the
  field surrounding Q1553+113 ($z_{\rm em} = 0.360$).
}
\label{fig:Q1553+113_img}
\end{figure}

\begin{figure}
%\epsscale{0.8}
%\plotone{Figures/Q1553+113_hist.ps}
%\includegraphics[width=6in]{f33.eps}
\caption{Same as for Figure~\ref{fig:Q0026+1259_hist} but for the
  field surrounding Q1553+113 ($z_{\rm em} = 0.360$).
}
\label{fig:Q1553+113_hist}
\end{figure}

%%%%%%%%%%%%%%%%%%%%%%%
% PKS2005-489
\begin{figure}
%\epsscale{0.8}
%\plotone{Figures/PKS2005-489_img.ps}
%\includegraphics[width=6in]{f34.eps}
\caption{Same as for Figure~\ref{fig:Q0026+1259_img} but for the
  field surrounding PKS2005--489 ($z_{\rm em} = 0.071$).
}
\label{fig:PKS2005-489_img}
\end{figure}

\begin{figure}
%\epsscale{0.8}
%\plotone{Figures/PKS2005-489_hist.ps}
%\includegraphics[width=6in]{f35.eps}
\caption{Same as for Figure~\ref{fig:Q0026+1259_hist} but for the
  field surrounding PKS2005--489 ($z_{\rm em} = 0.071$).
}
\label{fig:PKS2005-489_hist}
\end{figure}

%%%%%%%%%%%%%%%%%%%%%%%
% FJ2155-092
\begin{figure}
%\epsscale{0.8}
%\plotone{Figures/FJ2155-092_img.ps}
%\includegraphics[width=6in]{f36.eps}
\caption{Same as for Figure~\ref{fig:Q0026+1259_img} but for the
  field surrounding FJ2155--092 ($z_{\rm em} = 0.170$).
}
\label{fig:FJ2155-092_img}
\end{figure}

\begin{figure}
%\epsscale{0.8}
%\plotone{Figures/FJ2155-092_hist.ps}
%\includegraphics[width=6in]{f37.eps}
\caption{Same as for Figure~\ref{fig:Q0026+1259_hist} but for the
  field surrounding FJ2155--092 ($z_{\rm em} = 0.170$).
}
\label{fig:FJ2155-092_hist}
\end{figure}

%%%%%%%%%%%%%%%%%%%%%%%
% PKS2155-304
\begin{figure}
%\epsscale{0.8}
%\plotone{Figures/PKS2155-304_img.ps}
%\includegraphics[width=6in]{f38.eps}
\caption{Same as for Figure~\ref{fig:Q0026+1259_img} but for the
  field surrounding PKS2155--304 ($z_{\rm em} = 0.116$).
}
\label{fig:PKS2155-304_img}
\end{figure}

\begin{figure}
%\epsscale{0.8}
%\plotone{Figures/PKS2155-304_hist.ps}
%\includegraphics[width=6in]{f39.eps}
\caption{Same as for Figure~\ref{fig:Q0026+1259_hist} but for the
  field surrounding PKS2155--304 ($z_{\rm em} = 0.116$).
}
\label{fig:PKS2155-304_hist}
\end{figure}

\clearpage

%%%%%%%%%%%%%%%%%%%%%%%%%%%%%%%%%%%%%%%%%%%%%%%%%%%%%%%%%%%%
\begin{figure*}[ht]
%\epsscale{0.8}
%\plotone{Figures/fig_stargal.ps}
\includegraphics[height=6.5in,angle=90]{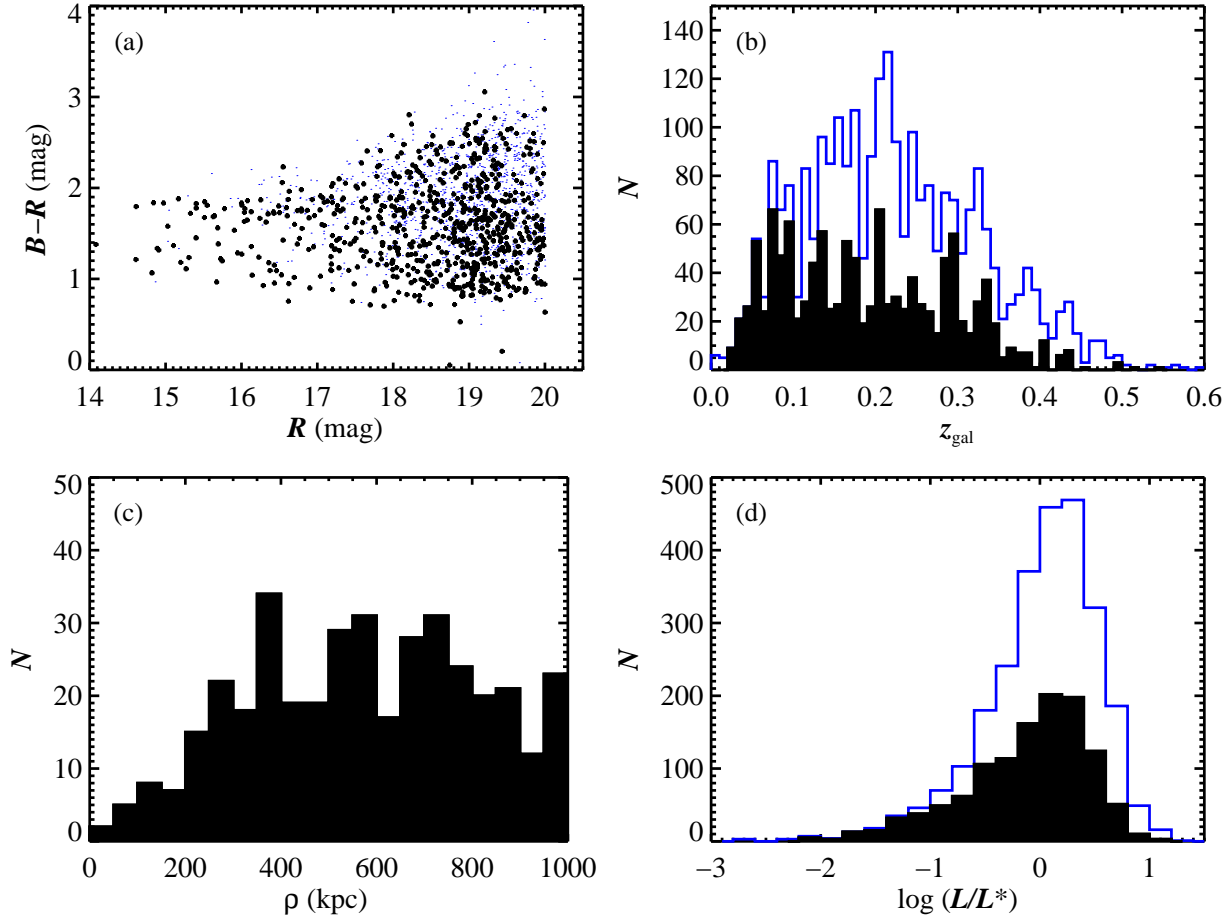}
\caption{(a) Color-magnitude diagram of all galaxies with
 redshifts $z>0.005$ (blue dots) in LCO/WFCCD survey and galaxies with 
 $0.02 < z < (z_{\rm em} - 0.01)$ (black symbols). 
(b) Histogram of the galaxy redshifts for sources detected in our
survey.  The open (blue) histogram shows all sources with $z>0.005$, while the
solid black histogram is restricted to $0.02 < z < (z_{\rm em} -
0.01)$ and therefore represents the sample that may be used to study
associations with the low $z$ IGM. 
(c) Distribution of impact parameters $\rho$ for the galaxies in our
survey with $0.02 < z < (z_{\rm em} - 0.01)$.   The incidence of
objects does not increase as $N \propto \rho$ for $\rho \gtrsim
400$\,kpc because of the (nearly) fixed angular extent of the survey.
(d) Luminosity distribution for the galaxies in our LCO/WFCCD survey
restricted to the sample of sources with $z>0.005$ (blue, open
histogram) and
$0.02 < z < (z_{\rm em} - 0.01)$ (black, solid histogram). 
}
\label{fig:summary}
\end{figure*}

\begin{figure}
%\epsscale{0.8}
%\plotone{Figures/fig_stargal.ps}
%\includegraphics[scale=0.6]{f41.eps}
\includegraphics[height=3.5in,angle=90]{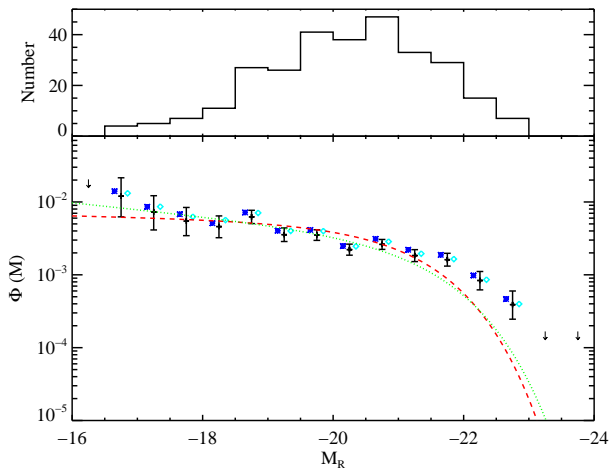}
\caption{Luminosity function estimated from the 17 fields of the
  LCO/WFCCD survey that have greater than $70\%$ completeness to $10'$ from the quasar to $R \le
  19.5$\,mag.  Black points with error bars show the results without a
  correction for incompleteness and assume Poisson uncertainties based on the
  number of galaxies detected (upper panel).  The blue stars and cyan
  diamonds show
  the estimated $\Phi(M)$ values after adopting magnitude-independent
  and dependent completeness corrections respectively (see the text
  for a full description). The dashed (red) and dotted (green) curves 
  show the luminosity functions derived from the SDSS by
  \cite{blanton03} and \cite{montero09} respectively, corrected
  to our assumed cosmology ($h=0.72$).  The offset at high luminosity may be the
  result of an Eddington bias in our estimation of $\Phi(M)$ and/or a
  modest offset between our $R$-band photometry and the SDSS $r$-band
  measurements.
}
\label{fig:lumfunc}
\end{figure}

\begin{figure}
%\epsscale{0.8}
%\plotone{Figures/fig_stargal.ps}
%\includegraphics[scale=0.6]{f42.eps}
\includegraphics[height=3.5in]{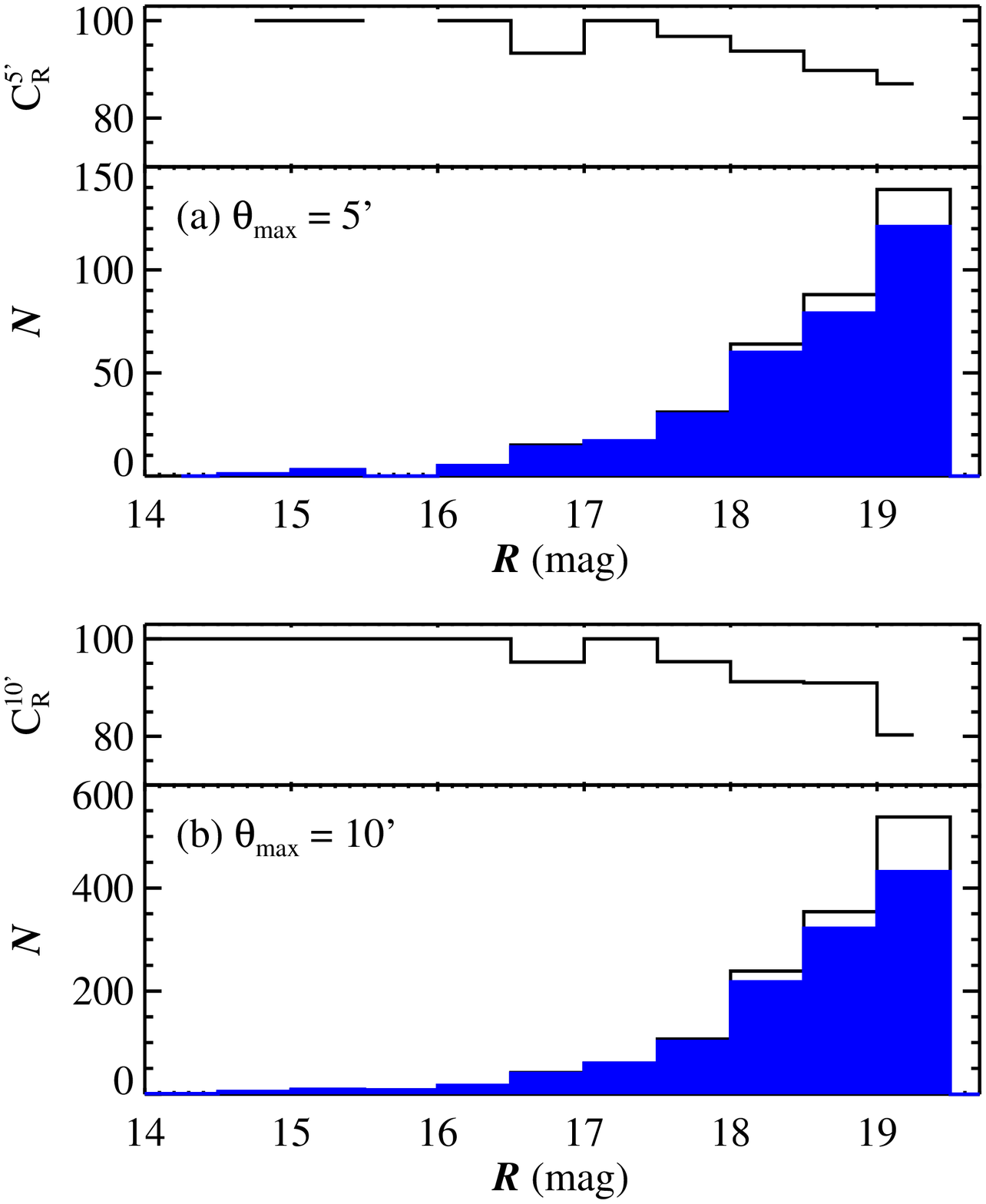}
\caption{The lower plots of each panel show histograms of the targeted (open,
  black) and observed (solid, blue) galaxies in the 17 fields used to
  estimate the luminosity function (Figure~\ref{fig:lumfunc}).  These
  histograms are shown for (a) $\theta_{\rm max} = 5'$ and (b)
  $\theta_{\rm max} = 10'$.  In the upper plot of each panel, we show
  the completeness percentile $C_R$ as a function of apparent
  magnitude.  These values are used in the magnitude-dependent
  completeness correction of the luminosity function.
}
\label{fig:mag_complete}
\end{figure}

%%%%%%%%%%%%%%%%%%%%%%%%%%%%%%%%%%%%%%%%%%%%%%%
\section{Summary}
\label{sec:summary}

This paper serves to define our LCO/WFCCD galaxy survey which was designed
to study the IGM/galaxy connection at $z \lesssim 0.2$.  We have
listed $R$-band photometry for objects in each field, identified the
objects which satisfy our targeting criteria, provide (online) the
extracted and coadded 1D spectra, and tabulate the redshifts 
for all sources with an unambiguous identification.
Table~\ref{tab:all_close} summarizes properties of the galaxies discovered in the 20
fields restricted to $0.005 < z < (z_{\rm em} + 0.01$).  
Again, we note that approximately half of the fields have been
surveyed for galaxies by other groups that our results should not be
considered a complete census. 
This table and the
redshift histograms reveal the types of galaxy overdensities that are
typical of other low $z$ galaxy surveys.  

Figure~\ref{fig:summary} presents a color-magnitude diagram for the
galaxies, and summarizes the redshift, luminosity, and impact
parameter distributions of the sample.  The set of galaxies with
$z<(z_{\rm em}-0.01$) is predominantly at redshifts $z=0.1$ -- 0.3, with
luminosities $L = 0.1$--$5 L^*$, and located at 
impact parameters of several hundred kpc from the quasar.   These distributions are a
complex convolution of the magnitude limit and angular extent of the
survey fields, and the emission redshift distribution of the targeted
quasars.  For example, the impact parameter distribution does not
scale as $N \propto \rho$ because the fixed angular extent of the
survey restricts galaxies to $\rho < 1$\,Mpc at $z < 0.1$.
Nevertheless, Figure~\ref{fig:summary} provides an overview
of this survey's utility for probing the low $z$ IGM.

To crudely assess the completeness and global characteristics of the
survey, we have also constructed an $R$-band luminosity function from the
dataset.  Specifically, we restricted the survey of each field to
$R_{\rm max} = 19.5$\,mag, $\theta_{\rm max} = 10'$, and 
$0.02 < z < min[(z_{\rm em} - 0.01), 0.2]$.  The maximum redshift was imposed to
minimize effects from having targeted fields with a known, luminous
quasar and also to facilitate comparison with the Sloan Digital Sky
Survey (SDSS).
We calculated the effective comoving volume 
$V_{\rm eff}$ for a given absolute magnitude $M$ and evaluated the
luminosity function,

\begin{equation}
\Phi(M) = \frac{n_{\rm gal}}{V_{\rm eff} \, \Delta M} \cmma
\end{equation}
within magnitude bins of $\Delta M = 0.5$\,mag.  The analysis was
further restricted to the 17 fields which have a completeness
percentile
$C_{19.5}^{10'} > 70\%$ (Table~\ref{tab:fields}).  
The raw evaluation of $\Phi(M)$ is shown as black points in
the lower panel of Figure~\ref{fig:lumfunc}, where the error bars reflect
only Poisson uncertainty ($1\sigma$ equivalent) in $n_{\rm gal}$.  Upper
limits correspond to 95$\%$\,c.l.  We have also performed a jack-knife
analysis of the survey and recover similar scatter in the measurements
albeit with a bias toward lower values.  

We have also attempted to correct for incompleteness in the survey,
using two approaches. The blue stars in the figure
include a completeness correction estimated by scaling $n_{\rm
  gal}$ in each field by its corresponding completeness, i.e.\
$1/C_{19.5}^{10'}$.  This increases $\Phi(M)$
in nearly every bin by $\approx 20\%$, i.e.\ the correction is
essentially independent of apparent magnitude.   As
Figure~\ref{fig:mag_complete} demonstrates, however, the survey has
higher incompleteness for fainter galaxies.  Therefore, we have used
the results given in Figure~\ref{fig:mag_complete} to increment the
contribution of each galaxy to $\Phi(M)$ according to each galaxy's
apparent magnitude.  These results are shown as the cyan diamonds in
Figure~\ref{fig:lumfunc}.

Overplotted on our evaluations
of $\Phi(M)$ are the luminosity functions derived for the $r$-band
from the Sloan Digital Sky Survey \citep{blanton03,montero09},
transformed to our assumed cosmology (e.g.\ $h=0.72$).  The agreement
between our evaluation and the SDSS estimations is remarkably good.
The only significant deviation is at the bright-end where we have
detected 2--$3\times$ more galaxies than predicted by the SDSS
luminosity functions. The offset could be the
result of an Eddington bias in our estimation of $\Phi(M)$ and/or a
modest offset between our $R$-band photometry and the SDSS $r$-band
(there is one notable for some of these galaxies which are also
observed by SDSS).  We also emphasize that the number of galaxies 
in these deviant bins is small ($\sim 20$; upper panel of
Figure~\ref{fig:lumfunc}). 
We conclude, therefore, that our combined survey provides a reasonably
representative sample of galaxies for low-$z$ IGM analysis.

In future papers (e.g.\ Prochaska et al.\ 2011, in prep), we 
explore the association of galaxies and their structures to the IGM
with particular focus on \ion{H}{1} \lya\ and \ion{O}{6} absorption.
We also encourage the acquisition of new, more sensitive UV spectral
datasets (i.e.\ with {\it HST}/COS) to further enhance this galaxy survey.

%%%%%%%%%%%%%%%%%%%%%%%%%%%%%%%%%%%%%%%%%%%%%%%
%%%%%%%%%%%%%%%%%%%%%%%%%%%%%%%%%%%%%%%%%%%%%%%
%%%%%%%%%%%%%%%%%%%%%%%%%%%%%%%%%%%%%%%%%%%%%%%

\acknowledgments

J. X. P. is partially supported by an NSF CAREER grant (AST--0548180).
We recognize the terrific staff at Las Campanas Observatory who were
instrumental in carrying out this program.  We kindly thank R. Weymann
who built the WFCCD spectrograph and supported this program.

%\bibliographystyle{/u/xavier/paper/Bibli/apj}
%\bibliography{/u/xavier/paper/Bibli/allrefs}

\begin{thebibliography}{51}
\expandafter\ifx\csname natexlab\endcsname\relax\def\natexlab#1{#1}\fi

\bibitem[{{Barton} \& {Cooke}(2009)}]{bc09}
{Barton}, E.~J., \& {Cooke}, J. 2009, \aj, 138, 1817

\bibitem[{{Bergeron} \& {Boisse}(1991)}]{bb91}
{Bergeron}, J., \& {Boisse}, P. 1991, Advances in Space Research, 11, 241

\bibitem[{{Bertin} \& {Arnouts}(1996)}]{bertin96}
{Bertin}, E., \& {Arnouts}, S. 1996, Astronomy and Astrophysics Supplement,
  117, 393

\bibitem[{{Blanton} {et~al.}(2003){Blanton}, {Hogg}, {Bahcall}, {Baldry},
  {Brinkmann}, {Csabai}, {Eisenstein}, {Fukugita}, {Gunn}, {Ivezi{\'c}},
  {Lamb}, {Lupton}, {Loveday}, {Munn}, {Nichol}, {Okamura}, {Schlegel},
  {Shimasaku}, {Strauss}, {Vogeley}, \& {Weinberg}}]{blanton03}
{Blanton}, M.~R., {et~al.} 2003, \apj, 594, 186

\bibitem[{{Bowen} {et~al.}(2001){Bowen}, {Jimenez}, {Jenkins}, \&
  {Pettini}}]{bjj+01}
{Bowen}, D.~V., {Jimenez}, R., {Jenkins}, E.~B., \& {Pettini}, M. 2001, \apj,
  547, 39

\bibitem[{{Bowen} {et~al.}(2002){Bowen}, {Pettini}, \& {Blades}}]{bowen+02}
{Bowen}, D.~V., {Pettini}, M., \& {Blades}, J.~C. 2002, \apj, 580, 169

\bibitem[{{Cagnoni} {et~al.}(2003){Cagnoni}, {Nicastro}, {Maraschi}, {Treves},
  \& {Tavecchio}}]{cagnoni+03}
{Cagnoni}, I., {Nicastro}, F., {Maraschi}, L., {Treves}, A., \& {Tavecchio}, F.
  2003, New A Rev., 47, 561

\bibitem[{{Cen} \& {Ostriker}(1999)}]{co99}
{Cen}, R., \& {Ostriker}, J.~P. 1999, \apj, 514, 1

\bibitem[{{Chen} {et~al.}(2010){Chen}, {Helsby}, {Gauthier}, {Shectman},
  {Thompson}, \& {Tinker}}]{chg+10}
{Chen}, H., {Helsby}, J.~E., {Gauthier}, J., {Shectman}, S.~A., {Thompson},
  I.~B., \& {Tinker}, J.~L. 2010, \apj, 714, 1521

\bibitem[{{Chen} \& {Mulchaey}(2009)}]{cm09}
{Chen}, H., \& {Mulchaey}, J.~S. 2009, \apj, 701, 1219

\bibitem[{{Chen} \& {Prochaska}(2000)}]{cp00}
{Chen}, H.-W., \& {Prochaska}, J.~X. 2000, \apjl, 543, L9

\bibitem[{{Chen} {et~al.}(2005){Chen}, {Prochaska}, {Weiner}, {Mulchaey}, \&
  {Williger}}]{cpw+05}
{Chen}, H.-W., {Prochaska}, J.~X., {Weiner}, B.~J., {Mulchaey}, J.~S., \&
  {Williger}, G.~M. 2005, \apjl, 629, L25

\bibitem[{{Cooksey} {et~al.}(2008){Cooksey}, {Prochaska}, {Chen}, {Mulchaey},
  \& {Weiner}}]{cpc+08}
{Cooksey}, K.~L., {Prochaska}, J.~X., {Chen}, H.-W., {Mulchaey}, J.~S., \&
  {Weiner}, B.~J. 2008, \apj, 676, 262

\bibitem[{{Danforth} \& {Shull}(2008)}]{ds08}
{Danforth}, C.~W., \& {Shull}, J.~M. 2008, \apj, 679, 194

\bibitem[{{Danforth} {et~al.}(2006){Danforth}, {Shull}, {Rosenberg}, \&
  {Stocke}}]{dsr+06}
{Danforth}, C.~W., {Shull}, J.~M., {Rosenberg}, J.~L., \& {Stocke}, J.~T. 2006,
  \apj, 640, 716

\bibitem[{{Dav{\'e}} {et~al.}(2001){Dav{\'e}}, {Cen}, {Ostriker}, {Bryan},
  {Hernquist}, {Katz}, {Weinberg}, {Norman}, \& {O'Shea}}]{daveetal01}
{Dav{\'e}}, R., {et~al.} 2001, \apj, 552, 473

\bibitem[{{Dunkley} {et~al.}(2009){Dunkley}, {Komatsu}, {Nolta}, {Spergel},
  {Larson}, {Hinshaw}, {Page}, {Bennett}, {Gold}, {Jarosik}, {Weiland},
  {Halpern}, {Hill}, {Kogut}, {Limon}, {Meyer}, {Tucker}, {Wollack}, \&
  {Wright}}]{wmap05}
{Dunkley}, J., {et~al.} 2009, \apjs, 180, 306

\bibitem[{{Fang} \& {Bryan}(2001)}]{fangandbryan01}
{Fang}, T., \& {Bryan}, G.~L. 2001, \apjl, 561, L31

\bibitem[{{Fang} {et~al.}(2002){Fang}, {Marshall}, {Lee}, {Davis}, \&
  {Canizares}}]{fangetal2002}
{Fang}, T., {Marshall}, H.~L., {Lee}, J.~C., {Davis}, D.~S., \& {Canizares},
  C.~R. 2002, \apjl, 572, L127

\bibitem[{{Howk} {et~al.}(2009){Howk}, {Ribaudo}, {Lehner}, {Prochaska}, \&
  {Chen}}]{howk+09}
{Howk}, J.~C., {Ribaudo}, J.~S., {Lehner}, N., {Prochaska}, J.~X., \& {Chen},
  H. 2009, \mnras, 396, 1875

\bibitem[{{Jenkins} {et~al.}(2005){Jenkins}, {Bowen}, {Tripp}, \&
  {Sembach}}]{jbt05}
{Jenkins}, E.~B., {Bowen}, D.~V., {Tripp}, T.~M., \& {Sembach}, K.~R. 2005,
  \apj, 623, 767

\bibitem[{{Landolt}(1992)}]{lan92}
{Landolt}, A.~U. 1992, \aj, 104, 372

\bibitem[{{Lanzetta} {et~al.}(1995){Lanzetta}, {Bowen}, {Tytler}, \&
  {Webb}}]{lbt+95}
{Lanzetta}, K.~M., {Bowen}, D.~V., {Tytler}, D., \& {Webb}, J.~K. 1995, \apj,
  442, 538

\bibitem[{{Lehner} {et~al.}(2009){Lehner}, {Prochaska}, {Kobulnicky},
  {Cooksey}, {Howk}, {Williger}, \& {Cales}}]{lpk+09}
{Lehner}, N., {Prochaska}, J.~X., {Kobulnicky}, H.~A., {Cooksey}, K.~L.,
  {Howk}, J.~C., {Williger}, G.~M., \& {Cales}, S.~L. 2009, \apj, 694, 734

\bibitem[{{Montero-Dorta} \& {Prada}(2009)}]{montero09}
{Montero-Dorta}, A.~D., \& {Prada}, F. 2009, \mnras, 399, 1106

\bibitem[{{Morris} {et~al.}(1993){Morris}, {Weymann}, {Dressler}, {McCarthy},
  {Smith}, {Terrile}, {Giovanelli}, \& {Irwin}}]{mwd+93}
{Morris}, S.~L., {Weymann}, R.~J., {Dressler}, A., {McCarthy}, P.~J., {Smith},
  B.~A., {Terrile}, R.~J., {Giovanelli}, R., \& {Irwin}, M. 1993, \apj, 419,
  524

\bibitem[{{Oppenheimer} \& {Dav{\'e}}(2009)}]{od09}
{Oppenheimer}, B.~D., \& {Dav{\'e}}, R. 2009, \mnras, 395, 1875

\bibitem[{{Penton} {et~al.}(2004){Penton}, {Stocke}, \& {Shull}}]{pss04}
{Penton}, S.~V., {Stocke}, J.~T., \& {Shull}, J.~M. 2004, \apjs, 152, 29

\bibitem[{{Prochaska} {et~al.}(2004){Prochaska}, {Chen}, {Howk}, {Weiner}, \&
  {Mulchaey}}]{pks0405_uv}
{Prochaska}, J.~X., {Chen}, H.-W., {Howk}, J.~C., {Weiner}, B.~J., \&
  {Mulchaey}, J. 2004, \apj, 617, 718

\bibitem[{{Prochaska} {et~al.}(2006){Prochaska}, {Weiner}, {Chen}, \&
  {Mulchaey}}]{pwc+06}
{Prochaska}, J.~X., {Weiner}, B.~J., {Chen}, H.-W., \& {Mulchaey}, J.~S. 2006,
  \apj, 643, 680

\bibitem[{{Savage} {et~al.}(2010){Savage}, {Narayanan}, {Wakker}, {Stocke},
  {Keeney}, {Shull}, {Sembach}, {Yao}, \& {Green}}]{snw+10}
{Savage}, B.~D., {et~al.} 2010, \apj, 719, 1526

\bibitem[{{Savage} {et~al.}(2002){Savage}, {Sembach}, {Tripp}, \&
  {Richter}}]{savage+02}
{Savage}, B.~D., {Sembach}, K.~R., {Tripp}, T.~M., \& {Richter}, P. 2002, \apj,
  564, 631

\bibitem[{{Scannapieco} {et~al.}(2006){Scannapieco}, {Pichon}, {Aracil},
  {Petitjean}, {Thacker}, {Pogosyan}, {Bergeron}, \& {Couchman}}]{spa+06}
{Scannapieco}, E., {Pichon}, C., {Aracil}, B., {Petitjean}, P., {Thacker},
  R.~J., {Pogosyan}, D., {Bergeron}, J., \& {Couchman}, H.~M.~P. 2006, \mnras,
  365, 615

\bibitem[{{Sembach} {et~al.}(2004){Sembach}, {Tripp}, {Savage}, \&
  {Richter}}]{sts+04}
{Sembach}, K.~R., {Tripp}, T.~M., {Savage}, B.~D., \& {Richter}, P. 2004,
  \apjs, 155, 351

\bibitem[{{Shone} {et~al.}(2010){Shone}, {Morris}, {Crighton}, \&
  {Wilman}}]{shone10}
{Shone}, A.~M., {Morris}, S.~L., {Crighton}, N., \& {Wilman}, R.~J. 2010,
  \mnras, 402, 2520

\bibitem[{{Shull} {et~al.}(1996){Shull}, {Stocke}, \& {Penton}}]{ssp96}
{Shull}, J.~M., {Stocke}, J.~T., \& {Penton}, S. 1996, \aj, 111, 72

\bibitem[{{Shull} {et~al.}(2003){Shull}, {Tumlinson}, \&
  {Giroux}}]{shulletal03}
{Shull}, J.~M., {Tumlinson}, J., \& {Giroux}, M.~L. 2003, \apjl, 594, L107

\bibitem[{{Steidel}(1993)}]{s93}
{Steidel}, C.~C. 1993, in Astrophysics and Space Science Library, Vol. 188, The
  Environment and Evolution of Galaxies, ed. J.~M. {Shull} \& H.~A. {Thronson},
  263--+

\bibitem[{{Stocke} {et~al.}(2006){Stocke}, {Penton}, {Danforth}, {Shull},
  {Tumlinson}, \& {McLin}}]{stockeetal06}
{Stocke}, J.~T., {Penton}, S.~V., {Danforth}, C.~W., {Shull}, J.~M.,
  {Tumlinson}, J., \& {McLin}, K.~M. 2006, \apj, 641, 217

\bibitem[{{Stocke} {et~al.}(1995){Stocke}, {Shull}, {Penton}, {Donahue}, \&
  {Carilli}}]{stocke+95}
{Stocke}, J.~T., {Shull}, J.~M., {Penton}, S., {Donahue}, M., \& {Carilli}, C.
  1995, \apj, 451, 24

\bibitem[{{Stoughton} {et~al.}(2002){Stoughton}, {Lupton}, {Bernardi},
  {Blanton}, {Burles}, {Castander}, {Connolly}, {Eisenstein}, {Frieman},
  {Hennessy}, {Hindsley}, {Ivezi{\'c}}, {Kent}, {Kunszt}, {Lee}, {Meiksin},
  {Munn}, {Newberg}, {Nichol}, {Nicinski}, {Pier}, {Richards}, {Richmond},
  {Schlegel}, {Smith}, {Strauss}, {SubbaRao}, {Szalay}, {Thakar}, {Tucker},
  {Vanden Berk}, {Yanny}, {Adelman}, {Anderson}, {Anderson}, {Annis},
  {Bahcall}, {Bakken}, {Bartelmann}, {Bastian}, {Bauer}, {Berman},
  {B{\"o}hringer}, {Boroski}, {Bracker}, {Briegel}, {Briggs}, {Brinkmann},
  {Brunner}, {Carey}, {Carr}, {Chen}, {Christian}, {Colestock}, {Crocker},
  {Csabai}, {Czarapata}, {Dalcanton}, {Davidsen}, {Davis}, {Dehnen},
  {Dodelson}, {Doi}, {Dombeck}, {Donahue}, {Ellman}, {Elms}, {Evans}, {Eyer},
  {Fan}, {Federwitz}, {Friedman}, {Fukugita}, {Gal}, {Gillespie}, {Glazebrook},
  {Gray}, {Grebel}, {Greenawalt}, {Greene}, {Gunn}, {de Haas}, {Haiman},
  {Haldeman}, {Hall}, {Hamabe}, {Hansen}, {Harris}, {Harris}, {Harvanek},
  {Hawley}, {Hayes}, {Heckman}, {Helmi}, {Henden}, {Hogan}, {Hogg}, {Holmgren},
  {Holtzman}, {Huang}, {Hull}, {Ichikawa}, {Ichikawa}, {Johnston}, {Kauffmann},
  {Kim}, {Kimball}, {Kinney}, {Klaene}, {Kleinman}, {Klypin}, {Knapp},
  {Korienek}, {Krolik}, {Kron}, {Krzesi{\'n}ski}, {Lamb}, {Leger},
  {Limmongkol}, {Lindenmeyer}, {Long}, {Loomis}, {Loveday}, {MacKinnon},
  {Mannery}, {Mantsch}, {Margon}, {McGehee}, {McKay}, {McLean}, {Menou},
  {Merelli}, {Mo}, {Monet}, {Nakamura}, {Narayanan}, {Nash}, {Neilsen},
  {Newman}, {Nitta}, {Odenkirchen}, {Okada}, {Okamura}, {Ostriker}, {Owen},
  {Pauls}, {Peoples}, {Peterson}, {Petravick}, {Pope}, {Pordes}, {Postman},
  {Prosapio}, {Quinn}, {Rechenmacher}, {Rivetta}, {Rix}, {Rockosi}, {Rosner},
  {Ruthmansdorfer}, {Sandford}, {Schneider}, {Scranton}, {Sekiguchi}, {Sergey},
  {Sheth}, {Shimasaku}, {Smee}, {Snedden}, {Stebbins}, {Stubbs}, {Szapudi},
  {Szkody}, {Szokoly}, {Tabachnik}, {Tsvetanov}, {Uomoto}, {Vogeley}, {Voges},
  {Waddell}, {Walterbos}, {Wang}, {Watanabe}, {Weinberg}, {White}, {White},
  {Wilhite}, {Wolfe}, {Yasuda}, {York}, {Zehavi}, \& {Zheng}}]{stoughton02}
{Stoughton}, C., {et~al.} 2002, \aj, 123, 485

\bibitem[{{Thom} \& {Chen}(2008)}]{tc08a}
{Thom}, C., \& {Chen}, H. 2008, \apj, 683, 22

\bibitem[{{Tripp} {et~al.}(2006){Tripp}, {Aracil}, {Bowen}, \&
  {Jenkins}}]{trippetal06}
{Tripp}, T.~M., {Aracil}, B., {Bowen}, D.~V., \& {Jenkins}, E.~B. 2006, \apjl,
  643, L77

\bibitem[{{Tripp} {et~al.}(2005){Tripp}, {Jenkins}, {Bowen}, {Prochaska},
  {Aracil}, \& {Ganguly}}]{tripp+05}
{Tripp}, T.~M., {Jenkins}, E.~B., {Bowen}, D.~V., {Prochaska}, J.~X., {Aracil},
  B., \& {Ganguly}, R. 2005, \apj, 619, 714

\bibitem[{{Tripp} {et~al.}(1998){Tripp}, {Lu}, \& {Savage}}]{tripp+98}
{Tripp}, T.~M., {Lu}, L., \& {Savage}, B.~D. 1998, \apj, 508, 200

\bibitem[{{Tripp} \& {Savage}(2000)}]{ts00}
{Tripp}, T.~M., \& {Savage}, B.~D. 2000, \apj, 542, 42

\bibitem[{{Tripp} {et~al.}(2008){Tripp}, {Sembach}, {Bowen}, {Savage},
  {Jenkins}, {Lehner}, \& {Richter}}]{tripp08}
{Tripp}, T.~M., {Sembach}, K.~R., {Bowen}, D.~V., {Savage}, B.~D., {Jenkins},
  E.~B., {Lehner}, N., \& {Richter}, P. 2008, \apjs, 177, 39

\bibitem[{{Tumlinson} {et~al.}(2005){Tumlinson}, {Shull}, {Giroux}, \&
  {Stocke}}]{tsg+05}
{Tumlinson}, J., {Shull}, J.~M., {Giroux}, M.~L., \& {Stocke}, J.~T. 2005,
  \apj, 620, 95

\bibitem[{{Wakker} \& {Savage}(2009)}]{wakker09}
{Wakker}, B.~P., \& {Savage}, B.~D. 2009, \apjs, 182, 378

\bibitem[{{Wiersma} {et~al.}(2010){Wiersma}, {Schaye}, {Dalla Vecchia},
  {Booth}, {Theuns}, \& {Aguirre}}]{wsd+10}
{Wiersma}, R.~P.~C., {Schaye}, J., {Dalla Vecchia}, C., {Booth}, C.~M.,
  {Theuns}, T., \& {Aguirre}, A. 2010, \mnras, 409, 132

\bibitem[{{Williger} {et~al.}(2006){Williger}, {Heap}, {Weymann}, {Dav{\'e}},
  {Ellingson}, {Carswell}, {Tripp}, \& {Jenkins}}]{willigeretal06}
{Williger}, G.~M., {Heap}, S.~R., {Weymann}, R.~J., {Dav{\'e}}, R.,
  {Ellingson}, E., {Carswell}, R.~F., {Tripp}, T.~M., \& {Jenkins}, E.~B. 2006,
  \apj, 636, 631

\end{thebibliography}

%\input{../../Tables/TONS180obj_sub.tab.tex}

\begin{deluxetable}{rrrcccccc}
\tablewidth{0pc}
\tablecaption{Ton~S~180: Object Summary \label{tab:TonS180}}
\tabletypesize{\footnotesize}
\tablehead{\colhead{ID} & \colhead{RA} & \colhead{DEC} 
& \colhead{$B$}
& \colhead{$R$} & \colhead{S/G$^a$} 
& \colhead{Area}& \colhead{flg$^b$} & \colhead{$z$} \\
 & & & (mag) & (mag) &  & ($\square''$) }
\startdata
1&00:57:27.9&--22:25:12&$ 9.86 \pm  0.08$&$ 9.37 \pm  0.06$&0.69& 253.5& 0& \nodata \\
2&00:56:52.8&--22:11:23&$12.07 \pm  0.08$&$11.42 \pm  0.06$&0.95&  40.1& 0& \nodata \\
3&00:56:34.2&--22:29:02&$15.35 \pm  0.09$&$14.34 \pm  0.06$&1.00&   3.5& 0& \nodata \\
4&00:56:33.8&--22:15:29&$19.34 \pm  0.10$&$17.53 \pm  0.06$&0.02&   5.8& 0& \nodata \\
5&00:56:32.1&--22:31:34&$21.96 \pm  0.20$&$20.94 \pm  0.10$&0.82&   3.3& 0& \nodata \\
6&00:56:32.0&--22:28:30&$19.97 \pm  0.11$&$18.10 \pm  0.06$&0.48&   4.5& 0& \nodata \\
7&00:56:33.0&--22:11:44&$19.09 \pm  0.10$&$17.16 \pm  0.06$&0.88&   5.0& 0& \nodata \\
8&00:56:32.4&--22:11:41&$23.32 \pm  0.54$&$21.59 \pm  0.13$&0.20&   4.3& 0& \nodata \\
9&00:56:32.0&--22:24:37&$21.65 \pm  0.18$&$20.28 \pm  0.08$&0.12&   3.9& 0& \nodata \\
10&00:56:32.6&--22:15:52&$21.40 \pm  0.15$&$20.02 \pm  0.07$&0.31&   2.9& 0& \nodata \\
\enddata
\tablenotetext{a}{Star/galaxy classifier calculated by SExtractor.
Values near unity indicate a stellar-like point-spread function.}
\tablenotetext{b}{This binary flag has the following code:
1: Survey target; 2: Spectrum taken; 4: Redshift measured.}
\tablecomments{[The complete version of this table is in the electronic edition of the Journal.  The printed edition contains only a sample.]}
\end{deluxetable}

\begin{deluxetable}{rrrcccccc}
\tablewidth{0pc}
\tablecaption{Ton~S~210: Object Summary \label{tab:TonS210}}
\tabletypesize{\footnotesize}
\tablehead{\colhead{ID} & \colhead{RA} & \colhead{DEC} 
& \colhead{$B$}
& \colhead{$R$} & \colhead{S/G$^a$} 
& \colhead{Area}& \colhead{flg$^b$} & \colhead{$z$} \\
 & & & (mag) & (mag) &  & ($\square''$) }
\startdata
1&01:21:01.7&--28:14:24&$16.10 \pm  0.10$&$13.78 \pm  0.06$&1.00&   4.7& 0& \nodata \\
2&01:21:00.1&--28:13:04&$18.81 \pm  0.09$&$17.36 \pm  0.07$&0.47&  10.4& 0& \nodata \\
3&01:20:58.4&--28:13:53&$19.05 \pm  0.09$&$17.87 \pm  0.07$&0.13&   4.9& 0& \nodata \\
4&01:20:56.6&--28:30:40&$21.65 \pm  0.23$&$18.85 \pm  0.07$&0.19&   4.9& 0& \nodata \\
5&01:20:57.4&--28:22:13&$19.54 \pm  0.10$&$18.31 \pm  0.07$&0.02&   6.9& 0& \nodata \\
6&01:20:57.7&--28:16:38&$19.50 \pm  0.09$&$18.63 \pm  0.07$&0.09&   5.1& 0& \nodata \\
7&01:20:57.2&--28:21:14&$20.68 \pm  0.12$&$18.19 \pm  0.07$&0.99&   2.1& 0& \nodata \\
8&01:20:57.1&--28:19:15&$24.85 \pm  1.70$&$21.36 \pm  0.12$&0.98&   3.1& 0& \nodata \\
9&01:20:56.6&--28:23:16&$22.62 \pm  0.29$&$19.87 \pm  0.07$&0.99&   1.9& 0& \nodata \\
10&01:20:55.9&--28:30:45&$22.82 \pm  0.43$&$20.35 \pm  0.09$&1.00&   4.3& 0& \nodata \\
\enddata
\tablenotetext{a}{Star/galaxy classifier calculated by SExtractor.
Values near unity indicate a stellar-like point-spread function.}
\tablenotetext{b}{This binary flag has the following code:
1: Survey target; 2: Spectrum taken; 4: Redshift measured.}
\tablecomments{[The complete version of this table is in the electronic edition of the Journal.  The printed edition contains only a sample.]}
\end{deluxetable}

\begin{deluxetable}{rrrccccc}
\tablewidth{0pc}
\tablecaption{PKS0312-770: Object Summary \label{tab:PKS0312-770}}
\tabletypesize{\footnotesize}
\tablehead{\colhead{ID} & \colhead{RA} & \colhead{DEC} 
& \colhead{$R$} & \colhead{S/G$^a$} 
& \colhead{Area}& \colhead{flg$^b$} & \colhead{$z$} \\
 & & & (mag) &  & ($\square''$) }
\startdata
1&03:11:42.7&--76:33:44&$12.65 \pm  0.01$&1.00&   7.2& 0& \nodata \\
2&03:12:28.6&--76:33:42&$18.71 \pm  0.02$&0.12&  10.2& 0& \nodata \\
3&03:12:10.2&--76:33:40&$17.71 \pm  0.01$&1.00&   4.6& 0& \nodata \\
4&03:14:33.7&--76:33:39&$16.63 \pm  0.01$&1.00&   4.6& 0& \nodata \\
5&03:12:34.0&--76:33:31&$19.25 \pm  0.03$&0.85&   3.8& 0& \nodata \\
6&03:11:25.3&--76:33:23&$20.89 \pm  0.08$&0.12&   7.8& 0& \nodata \\
7&03:12:52.6&--76:33:25&$19.44 \pm  0.03$&0.56&   5.2& 0& \nodata \\
8&03:10:19.9&--76:33:15&$20.60 \pm  0.10$&0.98&   2.7& 0& \nodata \\
9&03:13:01.0&--76:33:22&$21.29 \pm  0.12$&1.00&   5.5& 0& \nodata \\
10&03:14:36.7&--76:33:23&$20.19 \pm  0.05$&0.97&   8.1& 0& \nodata \\
\enddata
\tablenotetext{a}{Star/galaxy classifier calculated by SExtractor.
Values near unity indicate a stellar-like point-spread function.}
\tablenotetext{b}{This binary flag has the following code:
1: Survey target; 2: Spectrum taken; 4: Redshift measured.}
\tablecomments{[The complete version of this table is in the electronic edition of the Journal.  The printed edition contains only a sample.]}
\end{deluxetable}

\begin{deluxetable}{rrrcccccc}
\tablewidth{0pc}
\tablecaption{PKS0558-504: Object Summary \label{tab:PKS0558-504}}
\tabletypesize{\footnotesize}
\tablehead{\colhead{ID} & \colhead{RA} & \colhead{DEC} 
& \colhead{$B$}
& \colhead{$R$} & \colhead{S/G$^a$} 
& \colhead{Area}& \colhead{flg$^b$} & \colhead{$z$} \\
 & & & (mag) & (mag) &  & ($\square''$) }
\startdata
1&05:59:38.2&--50:18:33&$10.41 \pm  0.09$&$ 9.57 \pm  0.06$&0.69& 149.9& 0& \nodata \\
2&05:59:20.5&--50:17:37&$11.53 \pm  0.09$&$10.80 \pm  0.06$&0.73&  48.9& 0& \nodata \\
3&05:58:56.4&--50:27:28&$12.41 \pm  0.09$&$11.06 \pm  0.06$&0.75&  38.3& 0& \nodata \\
4&05:58:48.3&--50:17:10&$ 9.97 \pm  0.09$&$ 9.98 \pm  0.06$&0.69&  97.9& 0& \nodata \\
5&05:58:44.9&--50:24:12&$10.51 \pm  0.09$&$10.05 \pm  0.06$&0.69&  94.4& 0& \nodata \\
6&05:58:45.6&--50:25:13&$14.00 \pm  0.09$&$13.27 \pm  0.06$&1.00&   5.8& 0& \nodata \\
7&05:58:45.7&--50:16:04&$12.78 \pm  0.09$&$12.20 \pm  0.06$&1.00&  14.2& 0& \nodata \\
8&05:58:39.9&--50:26:41&$18.09 \pm  0.09$&$17.13 \pm  0.06$&1.00&   2.5& 0& \nodata \\
9&05:58:42.0&--50:26:33&$12.30 \pm  0.09$&$11.76 \pm  0.06$&0.98&  20.0& 0& \nodata \\
10&05:58:43.5&--50:22:24&$13.99 \pm  0.09$&$13.12 \pm  0.06$&1.00&   6.4& 0& \nodata \\
\enddata
\tablenotetext{a}{Star/galaxy classifier calculated by SExtractor.
Values near unity indicate a stellar-like point-spread function.}
\tablenotetext{b}{This binary flag has the following code:
1: Survey target; 2: Spectrum taken; 4: Redshift measured.}
\tablecomments{[The complete version of this table is in the electronic edition of the Journal.  The printed edition contains only a sample.]}
\end{deluxetable}

\begin{deluxetable}{rrrcccccc}
\tablewidth{0pc}
\tablecaption{PG1004+130: Object Summary \label{tab:PG1004+130}}
\tabletypesize{\footnotesize}
\tablehead{\colhead{ID} & \colhead{RA} & \colhead{DEC} 
& \colhead{$B$}
& \colhead{$R$} & \colhead{S/G$^a$} 
& \colhead{Area}& \colhead{flg$^b$} & \colhead{$z$} \\
 & & & (mag) & (mag) &  & ($\square''$) }
\startdata
1&10:06:47.3&+12:46:52&$ 8.19 \pm  0.00$&$ 8.74 \pm  0.00$&0.69& 217.7& 0& \nodata \\
2&10:06:57.9&+12:40:53&$10.58 \pm  0.00$&$11.13 \pm  0.00$&0.75&  30.0& 0& \nodata \\
3&10:06:40.6&+12:47:50&$21.44 \pm  0.24$&$21.99 \pm  0.24$&1.00&   2.4& 0& \nodata \\
4&10:06:39.5&+12:40:58&$19.82 \pm  0.06$&$20.37 \pm  0.06$&0.22&   3.7& 0& \nodata \\
5&10:06:38.8&+12:37:14&$17.74 \pm  0.01$&$18.29 \pm  0.01$&1.00&   2.6& 0& \nodata \\
6&10:06:40.1&+12:48:43&$21.92 \pm  0.30$&$22.46 \pm  0.30$&0.97&   7.6& 0& \nodata \\
7&10:06:40.3&+12:51:47&$17.97 \pm  0.01$&$18.51 \pm  0.01$&1.00&   2.5& 0& \nodata \\
8&10:06:39.4&+12:44:47&$19.73 \pm  0.07$&$20.28 \pm  0.07$&0.12&   5.7& 0& \nodata \\
9&10:06:39.5&+12:45:04&$21.60 \pm  0.29$&$22.15 \pm  0.29$&0.35&   9.2& 0& \nodata \\
10&10:06:38.5&+12:37:47&$19.38 \pm  0.05$&$19.93 \pm  0.05$&0.15&   3.6& 0& \nodata \\
\enddata
\tablenotetext{a}{Star/galaxy classifier calculated by SExtractor.
Values near unity indicate a stellar-like point-spread function.}
\tablenotetext{b}{This binary flag has the following code:
1: Survey target; 2: Spectrum taken; 4: Redshift measured.}
\tablecomments{[The complete version of this table is in the electronic edition of the Journal.  The printed edition contains only a sample.]}
\end{deluxetable}

\begin{deluxetable}{rrrcccccc}
\tablewidth{0pc}
\tablecaption{HE1029-140: Object Summary \label{tab:HE1029-140}}
\tabletypesize{\footnotesize}
\tablehead{\colhead{ID} & \colhead{RA} & \colhead{DEC} 
& \colhead{$B$}
& \colhead{$R$} & \colhead{S/G$^a$} 
& \colhead{Area}& \colhead{flg$^b$} & \colhead{$z$} \\
 & & & (mag) & (mag) &  & ($\square''$) }
\startdata
1&10:31:16.6&--14:21:26&$14.12 \pm  0.00$&$13.23 \pm  0.00$&1.00&   6.2& 0& \nodata \\
2&10:31:15.3&--14:14:52&$15.22 \pm  0.00$&$13.97 \pm  0.01$&1.00&   3.9& 0& \nodata \\
3&10:31:13.1&--14:15:55&$16.15 \pm  0.01$&$14.62 \pm  0.01$&1.00&   2.6& 0& \nodata \\
4&10:31:13.4&--14:11:51&$16.64 \pm  0.01$&$15.80 \pm  0.01$&1.00&   2.2& 0& \nodata \\
5&10:31:13.9&--14:06:06&$18.95 \pm  0.03$&$17.90 \pm  0.01$&0.87&   6.1& 7&$ 0.13066$\\
6&10:31:11.8&--14:21:00&$16.48 \pm  0.01$&$15.64 \pm  0.00$&1.00&   2.8& 0& \nodata \\
7&10:31:12.5&--14:16:26&$17.81 \pm  0.01$&$16.50 \pm  0.01$&1.00&   2.1& 0& \nodata \\
8&10:31:12.2&--14:13:53&$19.50 \pm  0.05$&$18.30 \pm  0.02$&0.28&   4.9& 7&$ 0.19151$\\
9&10:31:12.5&--14:11:37&$20.58 \pm  0.07$&$18.95 \pm  0.02$&0.98&   2.6& 7&$-0.00008$\\
10&10:31:12.6&--14:11:10&$22.97 \pm  0.45$&$20.40 \pm  0.05$&1.00&   2.0& 0& \nodata \\
\enddata
\tablenotetext{a}{Star/galaxy classifier calculated by SExtractor.
Values near unity indicate a stellar-like point-spread function.}
\tablenotetext{b}{This binary flag has the following code:
1: Survey target; 2: Spectrum taken; 4: Redshift measured.}
\tablecomments{[The complete version of this table is in the electronic edition of the Journal.  The printed edition contains only a sample.]}
\end{deluxetable}

\begin{deluxetable}{rrrcccccc}
\tablewidth{0pc}
\tablecaption{PG1116+215: Object Summary \label{tab:PG1116+215}}
\tabletypesize{\footnotesize}
\tablehead{\colhead{ID} & \colhead{RA} & \colhead{DEC} 
& \colhead{$B$}
& \colhead{$R$} & \colhead{S/G$^a$} 
& \colhead{Area}& \colhead{flg$^b$} & \colhead{$z$} \\
 & & & (mag) & (mag) &  & ($\square''$) }
\startdata
1&11:18:30.2&+21:17:08&$18.76 \pm  0.05$&$10.10 \pm  0.03$&0.98& 125.2& 0& \nodata \\
2&11:18:29.2&+21:14:00&$17.34 \pm  0.01$&$15.64 \pm  0.01$&0.07&   8.0& 7&$ 0.06038$\\
3&11:18:25.4&+21:08:33&$20.25 \pm  0.05$&$19.17 \pm  0.02$&0.59&   5.4& 1& \nodata \\
4&11:18:25.1&+21:18:15&$23.88 \pm  1.14$&$21.55 \pm  0.14$&0.22&  14.2& 0& \nodata \\
5&11:18:25.9&+21:18:06&$21.21 \pm  0.18$&$19.83 \pm  0.06$&0.03&  15.4& 0& \nodata \\
6&11:18:25.3&+21:16:13&$23.11 \pm  0.57$&$20.92 \pm  0.10$&0.08&  12.8& 0& \nodata \\
7&11:18:24.9&+21:16:10&$22.87 \pm  0.59$&$20.75 \pm  0.09$&0.03&  16.2& 0& \nodata \\
8&11:18:25.9&+21:16:25&$20.53 \pm  0.10$&$18.95 \pm  0.03$&0.03&  12.2& 0& \nodata \\
9&11:18:26.9&+21:27:28&$16.91 \pm  0.01$&$15.06 \pm  0.01$&1.00&   3.2& 0& \nodata \\
10&11:18:26.0&+21:27:25&$21.56 \pm  0.21$&$18.46 \pm  0.01$&0.99&   4.5& 1& \nodata \\
\enddata
\tablenotetext{a}{Star/galaxy classifier calculated by SExtractor.
Values near unity indicate a stellar-like point-spread function.}
\tablenotetext{b}{This binary flag has the following code:
1: Survey target; 2: Spectrum taken; 4: Redshift measured.}
\tablecomments{[The complete version of this table is in the electronic edition of the Journal.  The printed edition contains only a sample.]}
\end{deluxetable}

\begin{deluxetable}{rrrcccccc}
\tablewidth{0pc}
\tablecaption{PG1211+143: Object Summary \label{tab:PG1211+143}}
\tabletypesize{\footnotesize}
\tablehead{\colhead{ID} & \colhead{RA} & \colhead{DEC} 
& \colhead{$B$}
& \colhead{$R$} & \colhead{S/G$^a$} 
& \colhead{Area}& \colhead{flg$^b$} & \colhead{$z$} \\
 & & & (mag) & (mag) &  & ($\square''$) }
\startdata
1&12:14:00.7&+14:13:16&$12.30 \pm  0.09$&$10.94 \pm  0.07$&0.99&  28.4& 0& \nodata \\
2&12:13:51.9&+14:13:23&$16.04 \pm  0.09$&$14.81 \pm  0.07$&0.03&  35.2& 0& \nodata \\
3&12:13:58.3&+14:13:05&$14.94 \pm  0.09$&$13.93 \pm  0.07$&1.00&   5.9& 0& \nodata \\
4&12:14:35.3&+14:13:44&$25.76 \pm  2.12$&$22.05 \pm  0.12$&0.98&   2.5& 0& \nodata \\
5&12:14:29.6&+14:14:05&$19.90 \pm  0.10$&$17.77 \pm  0.07$&0.07&   6.9& 7&$ 0.15545$\\
6&12:13:42.4&+14:14:13&$21.00 \pm  0.11$&$19.93 \pm  0.08$&0.12&   6.9& 0& \nodata \\
7&12:14:45.0&+14:13:57&$21.25 \pm  0.12$&$19.80 \pm  0.07$&0.13&   4.8& 0& \nodata \\
8&12:14:28.2&+14:14:05&$20.27 \pm  0.10$&$18.88 \pm  0.07$&0.13&   7.1& 1& \nodata \\
9&12:14:40.8&+14:14:01&$19.92 \pm  0.10$&$18.00 \pm  0.07$&0.78&   7.2& 1& \nodata \\
10&12:14:40.9&+14:14:06&$23.36 \pm  0.32$&$21.58 \pm  0.11$&0.82&   4.4& 0& \nodata \\
\enddata
\tablenotetext{a}{Star/galaxy classifier calculated by SExtractor.
Values near unity indicate a stellar-like point-spread function.}
\tablenotetext{b}{This binary flag has the following code:
1: Survey target; 2: Spectrum taken; 4: Redshift measured.}
\tablecomments{[The complete version of this table is in the electronic edition of the Journal.  The printed edition contains only a sample.]}
\end{deluxetable}

\begin{deluxetable}{rrrcccccc}
\tablewidth{0pc}
\tablecaption{PG1216+069: Object Summary \label{tab:PG1216+069}}
\tabletypesize{\footnotesize}
\tablehead{\colhead{ID} & \colhead{RA} & \colhead{DEC} 
& \colhead{$B$}
& \colhead{$R$} & \colhead{S/G$^a$} 
& \colhead{Area}& \colhead{flg$^b$} & \colhead{$z$} \\
 & & & (mag) & (mag) &  & ($\square''$) }
\startdata
1&12:18:32.1&+06:32:29&$28.60 \pm 23.69$&$23.06 \pm  0.41$&1.00&   1.6& 0& \nodata \\
2&12:18:38.6&+06:42:29&$15.83 \pm  0.00$&$14.86 \pm  0.00$&0.05&  33.1& 7&$ 0.00667$\\
3&12:18:35.1&+06:29:41&$19.17 \pm  0.05$&$18.07 \pm  0.02$&0.03&  20.7& 0& \nodata \\
4&12:18:33.7&+06:28:08&$24.86 \pm  1.37$&$22.17 \pm  0.14$&0.96&   3.1& 0& \nodata \\
5&12:18:33.7&+06:29:06&$22.62 \pm  0.25$&$20.61 \pm  0.05$&1.00&   2.6& 0& \nodata \\
6&12:18:33.2&+06:40:52&$19.86 \pm  0.13$&$16.77 \pm  0.02$&0.03&  57.2& 0& \nodata \\
7&12:18:33.6&+06:46:39&$14.59 \pm  0.00$&$13.99 \pm  0.00$&1.00&   3.4& 0& \nodata \\
8&12:18:33.0&+06:34:06&$20.03 \pm  0.04$&$18.67 \pm  0.02$&0.99&   3.4& 0& \nodata \\
9&12:18:33.4&+06:38:45&$24.76 \pm  1.82$&$21.16 \pm  0.10$&0.12&   5.6& 0& \nodata \\
10&12:18:33.1&+06:40:30&$22.14 \pm  0.37$&$18.97 \pm  0.04$&0.03&  16.6& 0& \nodata \\
\enddata
\tablenotetext{a}{Star/galaxy classifier calculated by SExtractor.
Values near unity indicate a stellar-like point-spread function.}
\tablenotetext{b}{This binary flag has the following code:
1: Survey target; 2: Spectrum taken; 4: Redshift measured.}
\tablecomments{[The complete version of this table is in the electronic edition of the Journal.  The printed edition contains only a sample.]}
\end{deluxetable}

\begin{deluxetable}{rrrcccccc}
\tablewidth{0pc}
\tablecaption{3C273: Object Summary \label{tab:3C273}}
\tabletypesize{\footnotesize}
\tablehead{\colhead{ID} & \colhead{RA} & \colhead{DEC} 
& \colhead{$B$}
& \colhead{$R$} & \colhead{S/G$^a$} 
& \colhead{Area}& \colhead{flg$^b$} & \colhead{$z$} \\
 & & & (mag) & (mag) &  & ($\square''$) }
\startdata
1&12:28:23.7&+02:09:17&$26.82 \pm  5.73$&$23.58 \pm  0.25$&1.00&   3.5& 0& \nodata \\
2&12:28:22.2&+01:55:30&$23.45 \pm  0.35$&$21.95 \pm  0.07$&1.00&   2.1& 0& \nodata \\
3&12:28:22.2&+01:55:01&$22.74 \pm  0.23$&$21.31 \pm  0.05$&1.00&   2.1& 0& \nodata \\
4&12:28:22.1&+01:54:39&$23.44 \pm  0.35$&$22.18 \pm  0.08$&0.98&   1.7& 0& \nodata \\
5&12:28:21.8&+01:51:53&$23.13 \pm  0.31$&$21.53 \pm  0.05$&1.00&   1.9& 0& \nodata \\
6&12:28:23.7&+02:09:08&$24.55 \pm  1.24$&$22.21 \pm  0.11$&0.98&   3.4& 0& \nodata \\
7&12:28:23.0&+02:02:25&$24.95 \pm  1.82$&$22.59 \pm  0.16$&0.15&   2.9& 0& \nodata \\
8&12:28:23.1&+02:02:11&$28.23 \pm 14.61$&$23.70 \pm  0.32$&0.98&   1.3& 0& \nodata \\
9&12:28:23.8&+02:08:12&$25.41 \pm  1.28$&$24.00 \pm  0.35$&0.98&   1.5& 0& \nodata \\
10&12:28:23.5&+02:05:35&$24.12 \pm  0.86$&$22.17 \pm  0.13$&0.90&   5.6& 0& \nodata \\
\enddata
\tablenotetext{a}{Star/galaxy classifier calculated by SExtractor.
Values near unity indicate a stellar-like point-spread function.}
\tablenotetext{b}{This binary flag has the following code:
1: Survey target; 2: Spectrum taken; 4: Redshift measured.}
\tablecomments{[The complete version of this table is in the electronic edition of the Journal.  The printed edition contains only a sample.]}
\end{deluxetable}

\begin{deluxetable}{rrrcccccc}
\tablewidth{0pc}
\tablecaption{Q1230+095: Object Summary \label{tab:Q1230+095}}
\tabletypesize{\footnotesize}
\tablehead{\colhead{ID} & \colhead{RA} & \colhead{DEC} 
& \colhead{$B$}
& \colhead{$R$} & \colhead{S/G$^a$} 
& \colhead{Area}& \colhead{flg$^b$} & \colhead{$z$} \\
 & & & (mag) & (mag) &  & ($\square''$) }
\startdata
1&12:33:53.8&+09:31:53&$11.39 \pm  0.01$&$ 9.17 \pm  0.01$&0.69& 217.9& 0& \nodata \\
2&12:32:44.9&+09:39:38&$11.28 \pm  0.00$&$10.81 \pm  0.00$&0.72&  46.6& 0& \nodata \\
3&12:32:44.6&+09:22:05&$17.48 \pm  0.01$&$16.13 \pm  0.01$&0.01&   8.5& 7&$ 0.11526$\\
4&12:32:42.9&+09:20:40&$19.51 \pm  0.06$&$18.59 \pm  0.03$&0.03&  12.2& 1& \nodata \\
5&12:32:44.0&+09:30:56&$17.58 \pm  0.01$&$16.02 \pm  0.01$&1.00&   2.8& 0& \nodata \\
6&12:32:43.0&+09:24:56&$18.57 \pm  0.02$&$16.88 \pm  0.01$&1.00&   2.6& 0& \nodata \\
7&12:32:42.9&+09:27:53&$22.81 \pm  0.45$&$20.24 \pm  0.05$&0.57&   2.5& 0& \nodata \\
8&12:32:43.6&+09:35:56&$21.20 \pm  0.13$&$19.57 \pm  0.04$&0.18&   2.9& 0& \nodata \\
9&12:32:41.8&+09:20:56&$18.48 \pm  0.02$&$16.35 \pm  0.01$&1.00&   4.6& 0& \nodata \\
10&12:32:43.6&+09:36:08&$22.53 \pm  0.37$&$20.08 \pm  0.05$&0.40&   3.0& 0& \nodata \\
\enddata
\tablenotetext{a}{Star/galaxy classifier calculated by SExtractor.
Values near unity indicate a stellar-like point-spread function.}
\tablenotetext{b}{This binary flag has the following code:
1: Survey target; 2: Spectrum taken; 4: Redshift measured.}
\tablecomments{[The complete version of this table is in the electronic edition of the Journal.  The printed edition contains only a sample.]}
\end{deluxetable}

\begin{deluxetable}{rrrcccccc}
\tablewidth{0pc}
\tablecaption{PKS1302-102: Object Summary \label{tab:PKS1302-102}}
\tabletypesize{\footnotesize}
\tablehead{\colhead{ID} & \colhead{RA} & \colhead{DEC} 
& \colhead{$B$}
& \colhead{$R$} & \colhead{S/G$^a$} 
& \colhead{Area}& \colhead{flg$^b$} & \colhead{$z$} \\
 & & & (mag) & (mag) &  & ($\square''$) }
\startdata
1&13:05:43.7&--10:30:07&$10.28 \pm  0.00$&$ 8.93 \pm  0.01$&0.69& 334.3& 0& \nodata \\
2&13:04:57.9&--10:25:47&$13.20 \pm  0.00$&$12.36 \pm  0.00$&1.00&  13.8& 0& \nodata \\
3&13:04:57.8&--10:26:05&$17.48 \pm  0.02$&$15.15 \pm  0.01$&1.00&   3.2& 0& \nodata \\
4&13:04:56.0&--10:29:18&$20.06 \pm  0.11$&$18.11 \pm  0.02$&0.09&   6.8& 7&$ 0.27252$\\
5&13:04:54.6&--10:39:58&$18.85 \pm  0.04$&$17.89 \pm  0.01$&0.13&   4.6& 7&$ 0.04576$\\
6&13:04:54.9&--10:33:57&$17.45 \pm  0.01$&$15.92 \pm  0.01$&1.00&   2.3& 0& \nodata \\
7&13:04:56.1&--10:25:29&$23.12 \pm  0.78$&$20.75 \pm  0.08$&0.98&   1.9& 0& \nodata \\
8&13:04:55.0&--10:32:28&$21.53 \pm  0.25$&$19.65 \pm  0.04$&0.13&   3.7& 0& \nodata \\
9&13:04:56.0&--10:21:04&$20.36 \pm  0.11$&$18.71 \pm  0.02$&0.15&   3.6& 1& \nodata \\
10&13:04:55.1&--10:29:17&$19.09 \pm  0.03$&$18.13 \pm  0.01$&0.98&   3.4& 0& \nodata \\
\enddata
\tablenotetext{a}{Star/galaxy classifier calculated by SExtractor.
Values near unity indicate a stellar-like point-spread function.}
\tablenotetext{b}{This binary flag has the following code:
1: Survey target; 2: Spectrum taken; 4: Redshift measured.}
\tablecomments{[The complete version of this table is in the electronic edition of the Journal.  The printed edition contains only a sample.]}
\end{deluxetable}

\begin{deluxetable}{rrrcccccc}
\tablewidth{0pc}
\tablecaption{PG1307+085: Object Summary \label{tab:PG1307+085}}
\tabletypesize{\footnotesize}
\tablehead{\colhead{ID} & \colhead{RA} & \colhead{DEC} 
& \colhead{$B$}
& \colhead{$R$} & \colhead{S/G$^a$} 
& \colhead{Area}& \colhead{flg$^b$} & \colhead{$z$} \\
 & & & (mag) & (mag) &  & ($\square''$) }
\startdata
1&13:09:49.7&+08:11:05&$11.72 \pm  0.00$&$10.34 \pm  0.01$&0.69&  84.1& 0& \nodata \\
2&13:09:08.0&+08:09:16&$13.90 \pm  0.00$&$13.12 \pm  0.00$&1.00&   6.7& 0& \nodata \\
3&13:09:07.3&+08:09:35&$20.00 \pm  0.09$&$14.10 \pm  0.02$&1.00&   4.8& 0& \nodata \\
4&13:09:09.7&+08:27:13&$22.33 \pm  0.33$&$20.41 \pm  0.06$&0.13&   3.3& 0& \nodata \\
5&13:09:09.5&+08:27:09&$21.29 \pm  0.16$&$20.35 \pm  0.07$&0.15&   5.2& 0& \nodata \\
6&13:09:07.9&+08:11:41&$17.33 \pm  0.01$&$16.59 \pm  0.01$&1.00&   2.3& 0& \nodata \\
7&13:09:09.1&+08:23:48&$17.21 \pm  0.01$&$16.48 \pm  0.01$&1.00&   2.3& 0& \nodata \\
8&13:09:09.6&+08:28:56&$17.98 \pm  0.01$&$16.94 \pm  0.01$&1.00&   2.2& 0& \nodata \\
9&13:09:07.6&+08:09:48&$23.62 \pm  1.08$&$20.83 \pm  0.10$&0.20&   5.1& 0& \nodata \\
10&13:09:09.5&+08:27:53&$24.44 \pm  1.45$&$20.89 \pm  0.07$&0.99&   1.8& 0& \nodata \\
\enddata
\tablenotetext{a}{Star/galaxy classifier calculated by SExtractor.
Values near unity indicate a stellar-like point-spread function.}
\tablenotetext{b}{This binary flag has the following code:
1: Survey target; 2: Spectrum taken; 4: Redshift measured.}
\tablecomments{[The complete version of this table is in the electronic edition of the Journal.  The printed edition contains only a sample.]}
\end{deluxetable}

\begin{deluxetable}{rrrcccccc}
\tablewidth{0pc}
\tablecaption{MRK1383: Object Summary \label{tab:MRK1383}}
\tabletypesize{\footnotesize}
\tablehead{\colhead{ID} & \colhead{RA} & \colhead{DEC} 
& \colhead{$B$}
& \colhead{$R$} & \colhead{S/G$^a$} 
& \colhead{Area}& \colhead{flg$^b$} & \colhead{$z$} \\
 & & & (mag) & (mag) &  & ($\square''$) }
\startdata
1&14:29:00.6&+01:26:39&$10.44 \pm  0.00$&$ 9.62 \pm  0.00$&0.69& 140.4& 0& \nodata \\
2&14:28:57.4&+01:24:05&$11.17 \pm  0.00$&$10.52 \pm  0.00$&0.73&  60.6& 0& \nodata \\
3&14:28:29.6&+01:07:36&$12.80 \pm  0.01$&$11.16 \pm  0.01$&0.87&  34.0& 0& \nodata \\
4&14:28:21.9&+01:06:47&$18.66 \pm  0.04$&$17.19 \pm  0.01$&0.01&   6.3& 1& \nodata \\
5&14:28:21.3&+01:06:46&$23.57 \pm  0.82$&$21.66 \pm  0.13$&0.98&   2.4& 0& \nodata \\
6&14:28:24.2&+01:27:37&$20.25 \pm  0.09$&$18.02 \pm  0.01$&1.00&   2.4& 0& \nodata \\
7&14:28:24.2&+01:27:51&$22.63 \pm  0.54$&$20.90 \pm  0.10$&0.12&   2.9& 0& \nodata \\
8&14:28:23.2&+01:18:37&$22.68 \pm  0.57$&$19.94 \pm  0.05$&0.99&   2.5& 0& \nodata \\
9&14:28:23.7&+01:26:27&$22.09 \pm  0.30$&$20.53 \pm  0.07$&0.93&   2.5& 0& \nodata \\
10&14:28:24.0&+01:28:01&$22.26 \pm  0.37$&$19.38 \pm  0.03$&0.98&   2.1& 0& \nodata \\
\enddata
\tablenotetext{a}{Star/galaxy classifier calculated by SExtractor.
Values near unity indicate a stellar-like point-spread function.}
\tablenotetext{b}{This binary flag has the following code:
1: Survey target; 2: Spectrum taken; 4: Redshift measured.}
\tablecomments{[The complete version of this table is in the electronic edition of the Journal.  The printed edition contains only a sample.]}
\end{deluxetable}

\begin{deluxetable}{rrrcccccc}
\tablewidth{0pc}
\tablecaption{Q1553+113: Object Summary \label{tab:Q1553+113}}
\tabletypesize{\footnotesize}
\tablehead{\colhead{ID} & \colhead{RA} & \colhead{DEC} 
& \colhead{$B$}
& \colhead{$R$} & \colhead{S/G$^a$} 
& \colhead{Area}& \colhead{flg$^b$} & \colhead{$z$} \\
 & & & (mag) & (mag) &  & ($\square''$) }
\startdata
1&15:55:06.9&+11:14:54&$12.14 \pm  0.00$&$11.18 \pm  0.00$&0.96&  30.4& 0& \nodata \\
2&15:55:00.4&+10:59:45&$18.57 \pm  0.05$&$17.24 \pm  0.02$&0.17&  23.0& 0& \nodata \\
3&15:55:02.2&+10:59:33&$17.53 \pm  0.02$&$16.71 \pm  0.01$&0.03&  22.3& 0& \nodata \\
4&15:55:00.7&+11:04:26&$16.22 \pm  0.01$&$14.89 \pm  0.01$&0.02&  15.4& 7&$ 0.04052$\\
5&15:54:57.8&+11:01:35&$22.80 \pm  0.67$&$19.64 \pm  0.04$&0.20&   3.0& 1& \nodata \\
6&15:54:58.5&+11:19:32&$17.09 \pm  0.01$&$15.82 \pm  0.01$&1.00&   2.5& 0& \nodata \\
7&15:54:58.2&+11:17:38&$16.57 \pm  0.01$&$15.40 \pm  0.01$&1.00&   2.6& 0& \nodata \\
8&15:54:58.1&+11:21:23&$21.30 \pm  0.15$&$19.22 \pm  0.02$&0.99&   2.4& 0& \nodata \\
9&15:54:57.6&+11:21:27&$22.22 \pm  0.39$&$20.12 \pm  0.06$&0.12&   3.8& 0& \nodata \\
10&15:54:56.3&+11:05:18&$22.28 \pm  0.29$&$20.62 \pm  0.06$&0.98&   2.1& 0& \nodata \\
\enddata
\tablenotetext{a}{Star/galaxy classifier calculated by SExtractor.
Values near unity indicate a stellar-like point-spread function.}
\tablenotetext{b}{This binary flag has the following code:
1: Survey target; 2: Spectrum taken; 4: Redshift measured.}
\tablecomments{[The complete version of this table is in the electronic edition of the Journal.  The printed edition contains only a sample.]}
\end{deluxetable}

\begin{deluxetable}{rrrcccccc}
\tablewidth{0pc}
\tablecaption{PKS2005-489: Object Summary \label{tab:PKS2005-489}}
\tabletypesize{\footnotesize}
\tablehead{\colhead{ID} & \colhead{RA} & \colhead{DEC} 
& \colhead{$B$}
& \colhead{$R$} & \colhead{S/G$^a$} 
& \colhead{Area}& \colhead{flg$^b$} & \colhead{$z$} \\
 & & & (mag) & (mag) &  & ($\square''$) }
\startdata
1&20:09:57.0&--48:39:37&$24.07 \pm  0.55$&$22.75 \pm  0.14$&0.98&   3.0& 0& \nodata \\
2&20:10:27.4&--48:39:42&$25.30 \pm  9.99$&$23.33 \pm  0.22$&0.97&   1.7& 0& \nodata \\
3&20:10:18.0&--48:39:41&$25.25 \pm  9.99$&$23.47 \pm  0.26$&0.97&   3.7& 0& \nodata \\
4&20:10:11.1&--48:39:40&$25.20 \pm  2.48$&$20.77 \pm  0.07$&0.99&   2.5& 0& \nodata \\
5&20:10:17.0&--48:39:44&$24.47 \pm  0.97$&$22.50 \pm  0.14$&0.98&   3.9& 0& \nodata \\
6&20:09:45.5&--48:39:37&$21.46 \pm  0.13$&$20.35 \pm  0.07$&0.98&   2.6& 0& \nodata \\
7&20:09:10.9&--48:39:36&$26.00 \pm  2.54$&$23.67 \pm  0.24$&0.96&   1.7& 0& \nodata \\
8&20:09:19.3&--48:39:37&$24.66 \pm  0.93$&$23.31 \pm  0.22$&0.97&   2.6& 0& \nodata \\
9&20:09:23.9&--48:39:32&$22.53 \pm  0.33$&$20.48 \pm  0.07$&0.82&   3.7& 0& \nodata \\
10&20:10:22.2&--48:39:49&$23.53 \pm  0.46$&$22.03 \pm  0.11$&1.00&   2.5& 0& \nodata \\
\enddata
\tablenotetext{a}{Star/galaxy classifier calculated by SExtractor.
Values near unity indicate a stellar-like point-spread function.}
\tablenotetext{b}{This binary flag has the following code:
1: Survey target; 2: Spectrum taken; 4: Redshift measured.}
\tablecomments{[The complete version of this table is in the electronic edition of the Journal.  The printed edition contains only a sample.]}
\end{deluxetable}

\begin{deluxetable}{rrrcccccc}
\tablewidth{0pc}
\tablecaption{FJ2155-092: Object Summary \label{tab:FJ2155-092}}
\tabletypesize{\footnotesize}
\tablehead{\colhead{ID} & \colhead{RA} & \colhead{DEC} 
& \colhead{$B$}
& \colhead{$R$} & \colhead{S/G$^a$} 
& \colhead{Area}& \colhead{flg$^b$} & \colhead{$z$} \\
 & & & (mag) & (mag) &  & ($\square''$) }
\startdata
1&21:54:16.5&--09:20:36&$13.37 \pm  0.09$&$11.93 \pm  0.06$&1.00&  19.4& 0& \nodata \\
2&21:54:15.7&--09:29:17&$16.32 \pm  0.09$&$14.94 \pm  0.06$&0.68&   9.7& 0& \nodata \\
3&21:54:13.4&--09:14:10&$15.51 \pm  0.09$&$14.55 \pm  0.06$&1.00&   3.2& 0& \nodata \\
4&21:54:12.0&--09:23:15&$13.41 \pm  0.09$&$12.65 \pm  0.06$&1.00&  10.3& 0& \nodata \\
5&21:54:12.0&--09:12:53&$20.31 \pm  0.15$&$18.05 \pm  0.07$&0.22&   3.8& 0& \nodata \\
6&21:54:11.6&--09:12:54&$24.50 \pm  2.61$&$21.14 \pm  0.14$&0.13&   4.8& 0& \nodata \\
7&21:54:11.4&--09:20:01&$20.78 \pm  0.14$&$19.03 \pm  0.07$&0.99&   2.3& 0& \nodata \\
8&21:54:11.1&--09:18:56&$20.04 \pm  0.14$&$17.71 \pm  0.07$&0.22&   4.1& 0& \nodata \\
9&21:54:11.0&--09:19:03&$21.20 \pm  0.19$&$18.50 \pm  0.07$&0.99&   2.5& 0& \nodata \\
10&21:54:11.5&--09:15:46&$19.16 \pm  0.10$&$17.82 \pm  0.06$&1.00&   2.5& 0& \nodata \\
\enddata
\tablenotetext{a}{Star/galaxy classifier calculated by SExtractor.
Values near unity indicate a stellar-like point-spread function.}
\tablenotetext{b}{This binary flag has the following code:
1: Survey target; 2: Spectrum taken; 4: Redshift measured.}
\tablecomments{[The complete version of this table is in the electronic edition of the Journal.  The printed edition contains only a sample.]}
\end{deluxetable}

\begin{deluxetable}{rrrcccccc}
\tablewidth{0pc}
\tablecaption{PKS2155-304: Object Summary \label{tab:PKS2155-304}}
\tabletypesize{\footnotesize}
\tablehead{\colhead{ID} & \colhead{RA} & \colhead{DEC} 
& \colhead{$B$}
& \colhead{$R$} & \colhead{S/G$^a$} 
& \colhead{Area}& \colhead{flg$^b$} & \colhead{$z$} \\
 & & & (mag) & (mag) &  & ($\square''$) }
\startdata
1&21:59:04.0&--30:09:30&$ 9.80 \pm  0.08$&$ 9.27 \pm  0.06$&0.69& 265.2& 0& \nodata \\
2&21:58:12.6&--30:17:46&$12.89 \pm  0.08$&$12.00 \pm  0.06$&0.99&  21.0& 0& \nodata \\
3&21:58:11.4&--30:19:08&$15.88 \pm  0.09$&$14.82 \pm  0.06$&0.03&  22.0& 7&$ 0.04528$\\
4&21:58:08.2&--30:11:58&$13.55 \pm  0.09$&$12.33 \pm  0.06$&1.00&  15.6& 0& \nodata \\
5&21:58:02.3&--30:15:35&$15.22 \pm  0.08$&$14.49 \pm  0.06$&1.00&   3.1& 0& \nodata \\
6&21:58:03.0&--30:06:04&$19.50 \pm  0.12$&$17.12 \pm  0.06$&0.04&   6.3& 0& \nodata \\
7&21:58:00.3&--30:19:49&$20.61 \pm  0.14$&$18.30 \pm  0.06$&0.89&   3.7& 0& \nodata \\
8&21:58:01.8&--30:02:37&$21.55 \pm  0.20$&$19.17 \pm  0.07$&0.48&   4.0& 0& \nodata \\
9&21:58:00.8&--30:07:21&$21.61 \pm  0.18$&$19.21 \pm  0.07$&1.00&   2.4& 0& \nodata \\
10&21:57:59.4&--30:21:38&$19.98 \pm  0.09$&$18.80 \pm  0.06$&0.98&   2.1& 0& \nodata \\
\enddata
\tablenotetext{a}{Star/galaxy classifier calculated by SExtractor.
Values near unity indicate a stellar-like point-spread function.}
\tablenotetext{b}{This binary flag has the following code:
1: Survey target; 2: Spectrum taken; 4: Redshift measured.}
\tablecomments{[The complete version of this table is in the electronic edition of the Journal.  The printed edition contains only a sample.]}
\end{deluxetable}

%%%%%%%%%%%%%%%%%%%%%%%%%%%%%%%%%%%%%%%%%%%%%%%%%
%\input{../../Tables/Q0026+1259obj_sub.tab.tex}
%\input{../../Tables/TONS180obj_sub.tab.tex}
%\input{../../Tables/TONS210obj_sub.tab.tex}
%\input{../../Tables/PKS0312-770obj_sub.tab.tex}
%\input{../../Tables/PKS0558-504obj_sub.tab.tex}
%\input{../../Tables/PG1004+130obj_sub.tab.tex}
%\input{../../Tables/HE1029-140obj_sub.tab.tex}
%\input{../../Tables/PG1116+215obj_sub.tab.tex}
%
%\clearpage
%
%\input{../../Tables/PG1211+143obj_sub.tab.tex}
%\input{../../Tables/PG1216+069obj_sub.tab.tex}
%\input{../../Tables/3C273obj_sub.tab.tex}
%\input{../../Tables/Q1230+095obj_sub.tab.tex}
%\input{../../Tables/PKS1302-102obj_sub.tab.tex}
%\input{../../Tables/PG1307+085obj_sub.tab.tex}
%\input{../../Tables/MRK1383obj_sub.tab.tex}
%\input{../../Tables/Q1553+113obj_sub.tab.tex}
%
%\clearpage
%
%\input{../../Tables/PKS2005-489obj_sub.tab.tex}
%\input{../../Tables/FJ2155-092obj_sub.tab.tex}
%\input{../../Tables/PKS2155-304obj_sub.tab.tex}
%
%\input{../../Tables/table_all_close.sub.tex}

%%%%%%%%%%%%%%%%%%%%%%%%%%%%%%%%%%%%%%%%%%%%5
%%%%%%%%%%%%%%%%%%%%%%%%%%%%%%%%%%%%%%%%%%%%5
%% FIGURES

\end{document}